\documentclass[twocolumn,showpacs,preprint2,amsmath,amssymb]{revtex4}
\usepackage{graphicx}
\usepackage{dcolumn}
\usepackage{epsfig}
\usepackage{bm}

\begin{document}

\topmargin 0.10in
\title{Untangling supernova-neutrino oscillations with beta-beam data}

\author{N.~Jachowicz$^{1}$, G.C.~McLaughlin$^{2}$, and C. Volpe$^{3}$}
\email{natalie.jachowicz@UGent.be}
\email{Gail_McLaughlin@ncsu.edu}
\email{volpe@ipno.in2p3.fr} 
\affiliation{$^{1}$ Department of Subatomic and Radiation Physics,\\ Ghent University, \\Proeftuinstraat 86, \\ B-9000 Gent, Belgium.\\
$^{2}$ Department of Physics,\\ North Carolina State University, \\
Raleigh, North Carolina 27695-8202.\\$^{3}$ Institut de Physique Nucl\'eaire, F-91406 Orsay cedex, France.}

\begin{abstract}
Recently, we suggested that low-energy beta-beam neutrinos can be very useful for the study of supernova neutrino interactions.  In this paper, we examine the use of a such experiment for the analysis of a supernova neutrino signal.   Since supernova neutrinos are oscillating, it is very likely that the terrestrial spectrum of supernova neutrinos of a given flavor will  not be the same as the  energy distribution with which these neutrinos were  first emitted. We demonstrate the efficacy of the proposed  method for untangling multiple neutrino spectra.  This is an essential feature of any model aiming at gaining information about the supernova mechanism, probing proto-neutron star physics, and understanding supernova nucleosynthesis, such as the neutrino process and the r-process.   We also consider the efficacy of different experimental approaches including measurements at multiple beam energies and detector configurations.
 
\end{abstract}
\pacs{25.30Pt}

\maketitle

\section{Introduction}

Neutrinos are of fundamental importance during the late stages of the evolution of a massive star. Although the supernova problem is not yet fully understood, it is thought that neutrinos play a  crucial role in the core-collapse and subsequent explosion.

When the thermonuclear fuel in the center of the star is exhausted, the lack of elements left to burn and produce the pressure to maintain its hydrostatic equilibrium makes it impossible for the star to prevent the implosion of the core.  
Once the core's mass is too large, it becomes gravitationally unstable, and starts to collapse.  The gravitational binding energy that is liberated will be released in the form of neutrinos.
At first, there is a plethora of neutrinos that are generated  by the neutronization processes accompanying the gravitational collapse. These are free to escape from the star.  During later stages, the densities and temperatures in the center become so high that despite their small interaction cross sections, the neutrino diffusion time  exceeds the time scale of the implosion.  The neutrinos are trapped, and the equilibrium in the center is extended to  weak interactions.  The gravitational binding energy leaks out in the form of neutrinos, (which are produced in pairs as well as by electron capture), on a neutrino diffusion timescale of about ten seconds.  A small portion of this energy is deposited
in the material above the proto-neutron star.  New hydrodynamic scenarios \cite{Blondin:2005wz,Burrows:2006uh}, as well as the inclusions of energy released by nuclear burning of infalling outer shells \cite{Bruenn:2006ew}, can be helpful in
creating an explosion. Nevertheless, the neutrinos remain an essential energy source for the explosion see e.g. \cite{Janka:2006fh}. 


 Although these neutrinos are only weakly interacting, this enormous amount of particles and energy traveling through the different layers of the star is able to cause a transformation of the elements synthesized during the preceding thermonuclear burning processes in the life of the star in the neutrino-process \cite{Hartmann:1991tk,Heger:2003mm}.  In addition, it is thought that a neutrino driven wind occurs at late time, which may produce the r-process elements \cite{Woosley:1994ux}.  


 The outcome of the element synthesis in the neutrino driven wind, and
the prospects for obtaining an r-process are quite sensitive to the
 relative numbers of neutrons and protons in the wind.  The relative
 numbers of neutrons and protons in the wind are determined primarily by
 the neutrino interactions $\nu_e + n \rightarrow p + e^-$, and $\bar
 {\nu}_e + p \rightarrow n + e^+$.  Small changes in the spectra will
 translate into changes in the relative number of neutrons and protons
 and therefore determine the extent to which very heavy r-process nuclei
 can be formed \cite{Fuller:1995ih,McLaughlin:1997qi}. Furthermore, small
 changes in the abundance distribution can be affected by  neutrino-
 nucleus interactions during and at the end of the rapid neutron capture
 process.  Because the thresholds for these reactions can be quite high,
 they are very sensitive to the neutrino spectra \cite{Haxton:1996ms,McLaughlin:1996eq,Terasawa:2004ju}.  The same considerations come into
 play for gamma ray bursts  \cite{Surman:2005kf,Surman:2003qt,Pruet:2003yn}.
 
 The neutrino-process in contrast occurs in the outer layers of the star,
 where neutrinos scatter on pre-existing nuclei. Certain rare nuclei can
 be formed by spallation processes~:  a neutrino scatters on a nucleus
 leaving an excited state within the nucleus that then decays by the
 emission of neutrons and/or protons.  Therefore it is not the capture of
 neutrinos on free nucleons, but the capture of neutrinos on nuclei that
 is the determine factor for this process.  Since thresholds are  high,
 changes in the spectra due to neutrino oscillations can have a dramatic
 effect on the abundances produced in these processes \cite
 {Yoshida:2006sk}.

As a matter of fact, neutrinos from the next galactic core collapse supernova are much anticipated, since they are the only particles giving us the chance to see deep into  the interior of the event and obtain information  about the processes driving the explosion and the influence of neutrinos on the events.
 Terrestrial supernova-neutrino detectors aim at the observation of supernova neutrinos through a variety of processes. The need for neutrino-nucleus cross sections comes into play.  Despite the small cross sections, neutrino-nucleus interactions are a powerful filter for information, owing to their energy, flavor, and spin sensitivity \cite{klaasprl,klaasprc}.

Currently on line detectors, such as Superkamiokande, MiniBoone,
and KamLAND are capable of detecting neutrinos from a galactic supernova,
for a review see e.g. \cite{Scholberg:2007nu}.
For a supernova 10 kiloparsecs away from the earth, a heavy water detector, such as 
the one which operated at the Sudbury Neutrino
Observatory would see several hundred events in all channels, 
$ {\nu}_e + d \rightarrow p + p + e^- $, 
$\overline{\nu}_e + d  \rightarrow n + n + e^- $,
 and $\nu + d \rightarrow n + p + \nu$ \cite{Beacom:1998yb}. 
Superkamiokande will record thousands
of events from the reaction $\overline{\nu}_e + p \rightarrow n + e^+ $
\cite{Beacom:1998ya}. In
addition there will be neutral current events on Oxygen-16, 
$\nu + ^{16}{O} \rightarrow \nu + ^{15}{0} + n +\gamma $, and
$\nu + ^{16}{O} \rightarrow \nu + ^{15}{N} + p +\gamma $ which are
detectable by way of the gammas e.g. \cite{Langanke:1995he, Kolbe:2003ys}.
In a proposed lead detector, such as OMNIS or LAND \cite{omnis,land}, the main signals will 
come from $\nu_e + ^{208}{Pb} \rightarrow ^{208}{Bi} + e^-$ and
$\nu + ^{208}{Pb} \rightarrow ^{208}{Pb} + \nu$, and analogue reactions on $^{56}$Fe.  
In these reactions the daughter nuclei
will typically spall one or two nucleons, providing information about the energy of the incoming neutrinos.  In KamLAND and MiniBoone there will
be an inverse beta-decay signal on protons, as well as a few events from
neutrino interactions with $^{12}$C. 

There are several uncertainties involved in any future measurement of supernova
neutrinos.  These include uncertainties in the range of 
predicted luminosities and spectra of the neutrinos, uncertainties in the type and degree of neutrino oscillations that occur, and the uncertainty in the
calculation of the detector neutrino-nucleus 
cross sections. 
Therefore, it is essential to detect neutrinos in all channels : both the neutral current
which is sensitive to all flavors 
and the two charged-current channels which are sensitive to $\nu_e$ and
$\bar{\nu}_e$.  When the neutrinos are originally emitted, the
$\nu_\mu$, $\bar{\nu}_\mu$, $\nu_\tau$ and $\bar{\nu}_\tau$ have the highest
energies, followed by the $\bar{\nu}_e$, and then the $\nu_e$.  Matter-enhanced neutrino transformation will mix these spectra  differently in the cases of  neutrinos and  antineutrinos.  
Hence, it is important to be sensitive to the different  neutrino flavors.
This provides the only way to maximize the information that can be obtained
about the original spectra.   

The type of neutrino flavor transformation that will occur in the supernova is a rapidly
developing field.  Effort is now being directed at understanding fully the neutrino
background terms \cite{Duan:2006jv,Sawyer:2005jk,Hannestad:2006nj,Duan:2007bt,Duan:2007mv,EstebanPretel:2007ec,Raffelt:2007cb,Raffelt:2007yz,Fogli:2007bk}
and their effect on understanding the supernova.
Furthermore, even if these terms are not important, phase effects due to multiple resonances come into
play \cite{Kneller:2005hf,Kneller:2007kg,Dasgupta:2005wn}.  This greatly complicates the simple picture of a single ``H'' and ``L'' 
resonance \cite{Dighe:1999bi,Engel:2002hg}.   Due to uncertainties in neutrino parameters and in supernova conditions, it is not yet clear how the 
neutrinos will transform during the relatively early times of the supernova, when most of the neutrinos 
are being released.  However, in addition to the simplest scenarios of two resonance points in the outer layers of the
star, realistic possibilities  include a complete  mixing of all flavors \cite{Sawyer:2005jk}, a 
partial oscillation between flavors \cite{Duan:2006jv} also called spectral swapping \cite{Raffelt:2007cb}, 
or a finely grained energy-dependent effect 
\cite{Kneller:2005hf,Kneller:2007kg}.  
An important  part of any future observation of 
supernova neutrinos will be to understand this physics.

The understanding of any astrophysical neutrino-nucleus interaction is limited by our understanding of  its cross section.
While the inverse beta decay on
protons is well understood, there is little data for  neutrino-nucleus cross
sections.  The exception is neutrino-nucleus cross section data
for $^{12}{\rm C}$ \cite{Drexlin,Albert,Kleinfeller,Imlay} 
measured using muon decay at rest (DAR) or
decay in flight spectra (DIF) spectra where measurements
have been reported to $\sim 10\%$ accuracy. Still, while the exclusive cross sections are well understood theoretically,
the predictions still fail to reproduce some of the inclusive ones
(see e.g. \cite{Cvolpe,Ckolbe1,Ckolbe2,Cikke}).  Electron neutrinos
from DAR are in the same energy regime as supernova neutrinos making
measurements at the 10\% level, such as those at the proposed $\nu$-SNS facility \cite{nuSNS}
attractive from the point of view of calibrating 
nuclear structure calculations. For a few other nuclei, such as Iron,
measurements have been reported at the $\sim 40\%$  level \cite{Berge:1989hr}.
At present, for most 
neutrino-nucleus cross sections, predictions rely on shell-model or RPA calculations.  In some cases one can use other measurements
as a guide for  the calculation. For example
charge-exchange data can show the location of the
Gamow-Teller resonance in the charged-current channel, 
as can the isovector part of the M1 strength in the neutral-current channel. Supernova
neutrinos however are expected to excite also first forbidden transitions
within the nucleus, for which it is more difficult to obtain useful data from
other sources with which to calibrate the calculations.

A recurring issue in the extraction of information from neutrino cross section measurements is the absence of monochromatic neutrino beams.  To some extent, this problem can be met by the flexibility offered by beta-beam neutrinos.
 The idea of producing neutrino beams from the beta-decay of boosted radioactive ions, stored in a storage ring, was first proposed by Zucchelli \cite{Zucchelli:sa}. The main goal of the proposed beta-beam facility was to explore the possible existence of CP violation in the lepton sector through the comparison of neutrino versus anti-neutrino oscillations. This intriguing proposal has triggered a feasibility study which is now ongoing \cite{cernweb,Lindroos:2003kp,Mezzetto:2003ub}. The sensitivity on the CP violating phase and on the third neutrino mixing angle that can be achieved with the original as well as other - higher energy - scenarios  is now being explored in great detail (see e.g. \cite{Campagne:2006yx}, 
\cite{Burguet-Castell:2003vv} and \cite{Volpe:2006in} for a review). The energy range interesting for core-collapse supernova physics could be covered by low energy beta-beams \cite{Volpe:2003fi}.
Several applications of such a facility have been studied recently going from nuclear structure studies
\cite{Volpe:2003fi,Serreau:2004kx,McLaughlin:2004va,Lazauskas:2007bs},
the study of fundamental interactions \cite{McLaughlin:2003yg,Volpe:2005iy,Balantekin:2005md,Balantekin:2006ga,Bueno:2006yq,Barranco:2007tz} and core-collapse supernova physics \cite{Volpe:2003fi,Jachowicz:2006xx}.

These applications could exploit either a devoted storage ring \cite{Serreau:2004kx}, or
low energy neutrino beams at off-axis from the standard storage ring \cite{Lazauskas:2007va}.
The application discussed here particularly benefits from very low ion boosts (i.e. $\gamma=3,6$).
Instead of running at such boosts, the very low energy neutrino fluxes could 
be obtained either by taking the flux at different parts of a detector
at low energy beta-beams \cite{Amanik:2007zy} or with an off-axis detector from the standard storage ring \cite{Lazauskas:2007va}.

In this paper, we examine what can be learned about the Galactic supernova neutrinos from the detector signal they produce.
We follow the approach proposed in \cite{Jachowicz:2006xx}, where  linear combinations of beta-beam spectra were shown to be able to mimic supernova-neutrino energy-distributions very accurately, thus allowing one to avoid the problems related to uncertainties and model dependencies in theoretical studies, and to the  lack of monochromatic neutrino beams in experiments.  As part of our analysis, we explicitly show how it is possible to disentangle a supernova-neutrino spectrum which has become ``mixed'' due to neutrino oscillations.  This can be done through a series of carefully chosen beta-beam spectral measurements.

The paper is organized as follows : After an introduction of the most common descriptions of the supernova neutrino spectra and the experimental neutrino energy distributions,  our technique for the construction of synthetic spectra is discussed.  The quality of the generated spectra is evaluated and we suggest  some opportunities for improvement of the technique   in actual implementations. Finally, section \ref{oscsec} elaborates on the information about the supernova and about oscillations  that can be extracted from the neutrino signal in a terrestrial detector.

\section{Neutrino Spectra}

\begin{figure*}
\vspace*{12.5cm}
\special{hscale=36 vscale=36 hsize=1500 vsize=600
         hoffset=-20 voffset=380 angle=-90 psfile="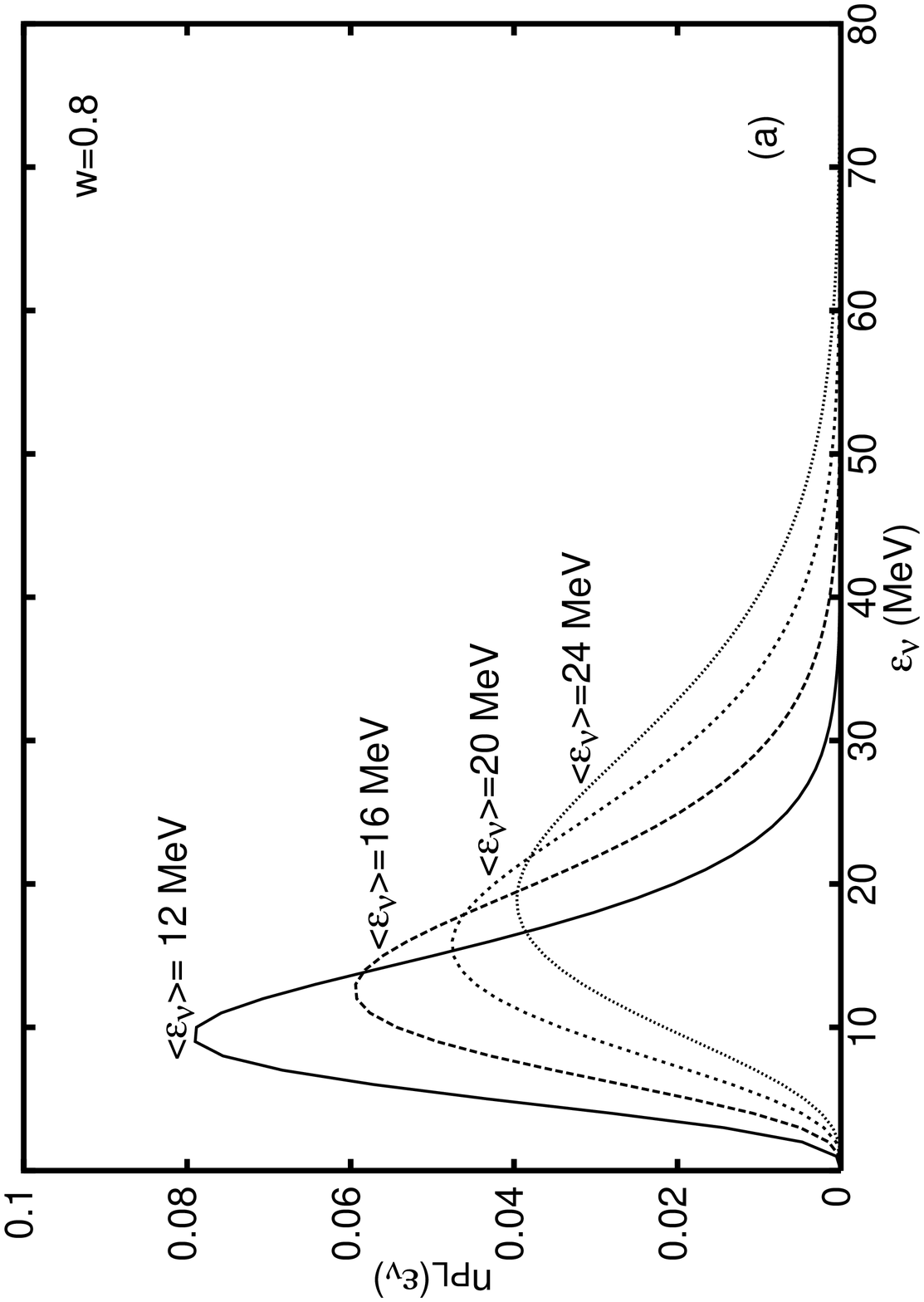"}
\special{hscale=36 vscale=36 hsize=1500 vsize=600
         hoffset=235 voffset=380 angle=-90 psfile="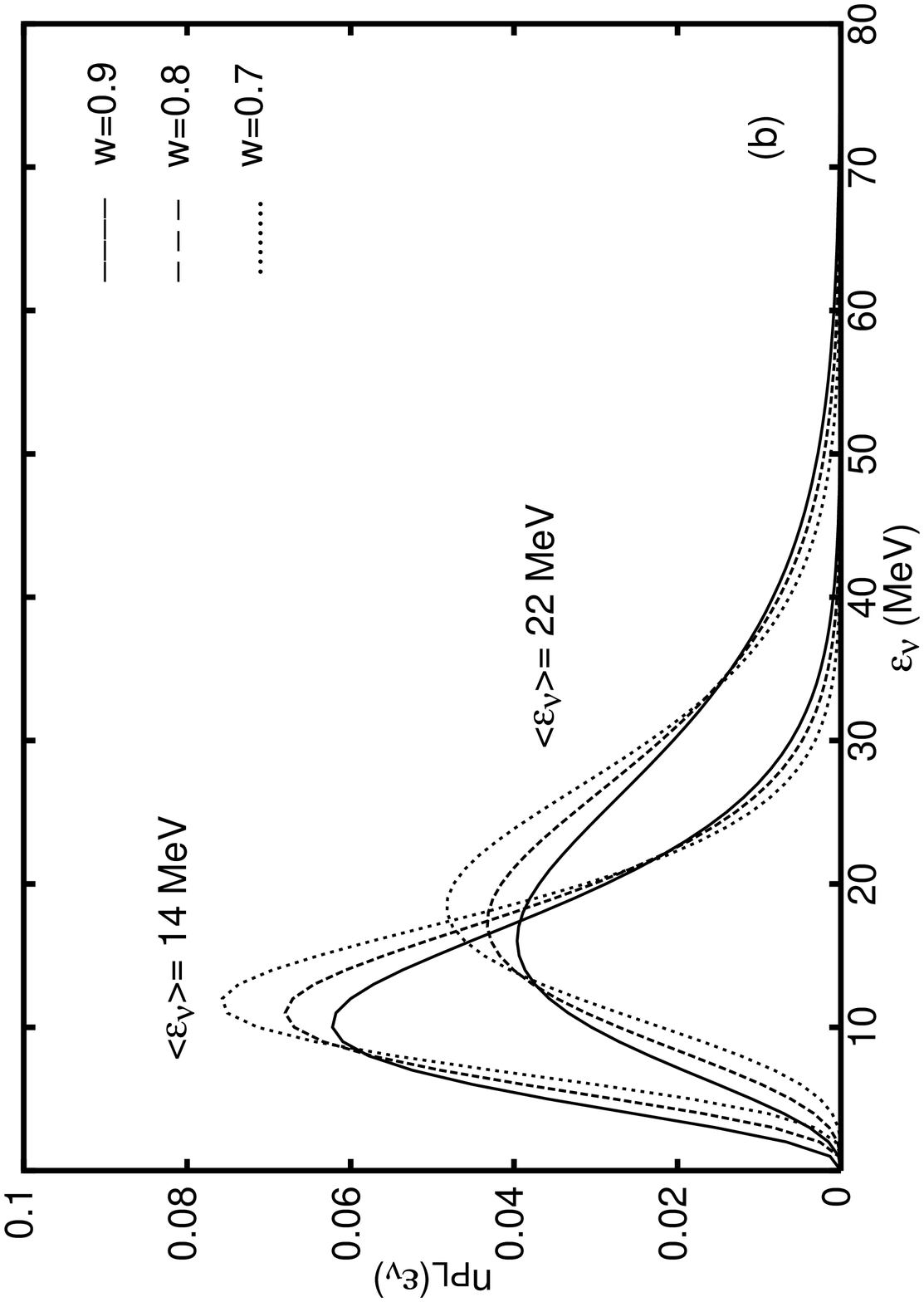"}
\special{hscale=36 vscale=36 hsize=1500 vsize=600
         hoffset=-20 voffset=200 angle=-90 psfile="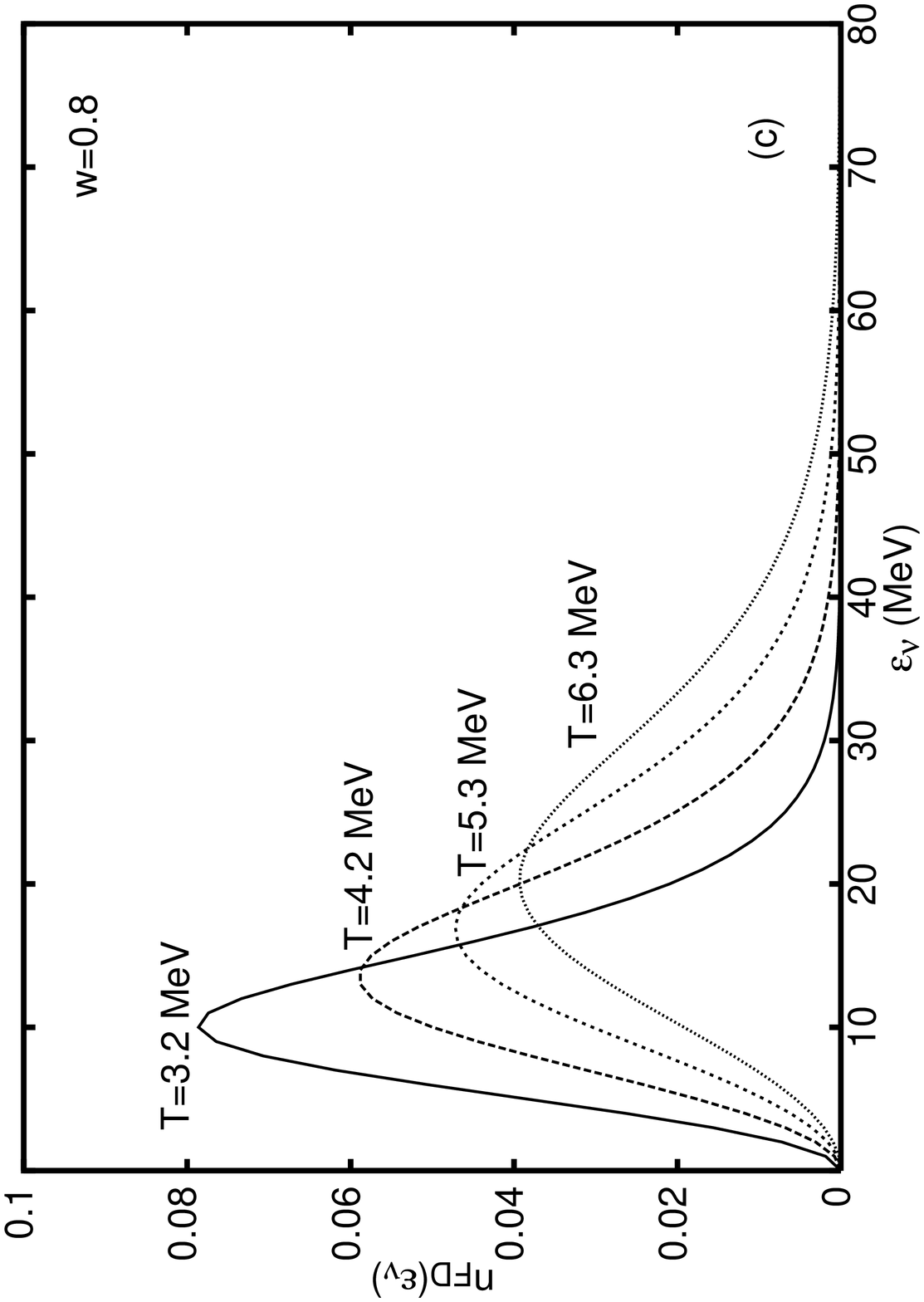"}
\special{hscale=36 vscale=36 hsize=1500 vsize=600
         hoffset=235 voffset=200 angle=-90 psfile="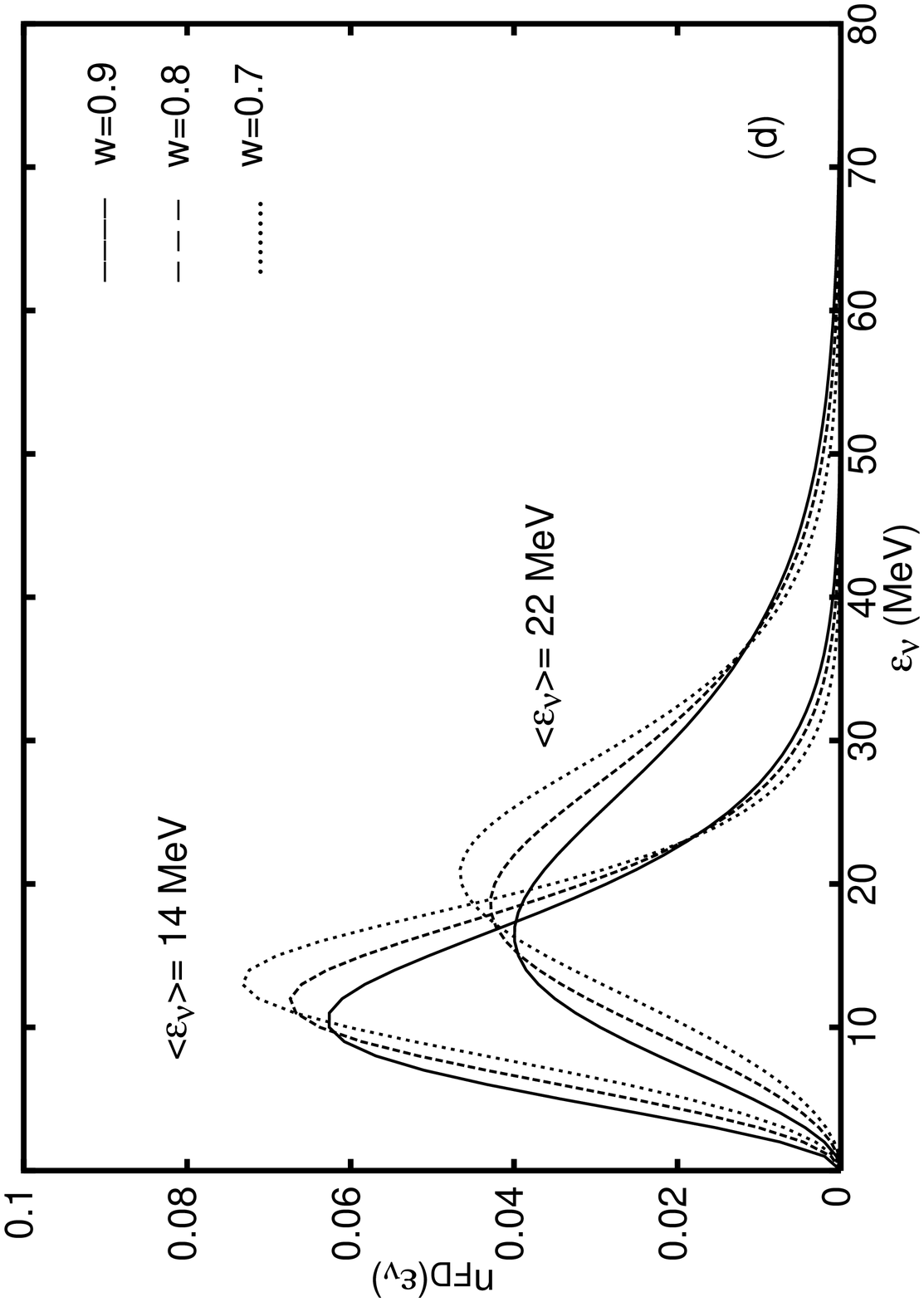"}
\caption{Neutrino energy-spectra.  The upper panels show different power-law parameterizations for supernova neutrinos.  Left, fixed width and varying average energies (a), right, varying width for two different values of the average energy (b). The lower panels show the equivalent Fermi-Dirac distributions. The width $w$ is expressed in units $w_0=\frac{\langle\varepsilon_{\nu}\rangle}{\sqrt{3}}$, for which power-law (with $\alpha$=2) and Fermi-Dirac spectra  ($\eta=\infty$) coincide.} 
\label{spec1}
\end{figure*}

\begin{figure}
\vspace*{6.5cm}
\special{hscale=36 vscale=36 hsize=1500 vsize=600
         hoffset=-20 voffset=200 angle=-90 psfile="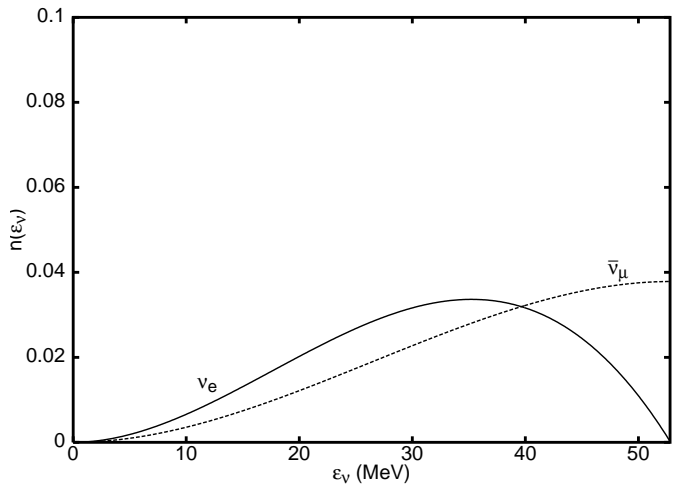"}
\caption{Michel spectra,  showing the $\nu_e$ and $\overline{\nu}_{\mu}$ energy distributions stemming from pion decay-at-rest.} 
\label{spec2}
\end{figure}

\begin{figure}
\vspace*{6.5cm}
\special{hscale=36 vscale=36 hsize=1500 vsize=600
         hoffset=-20 voffset=200 angle=-90 psfile="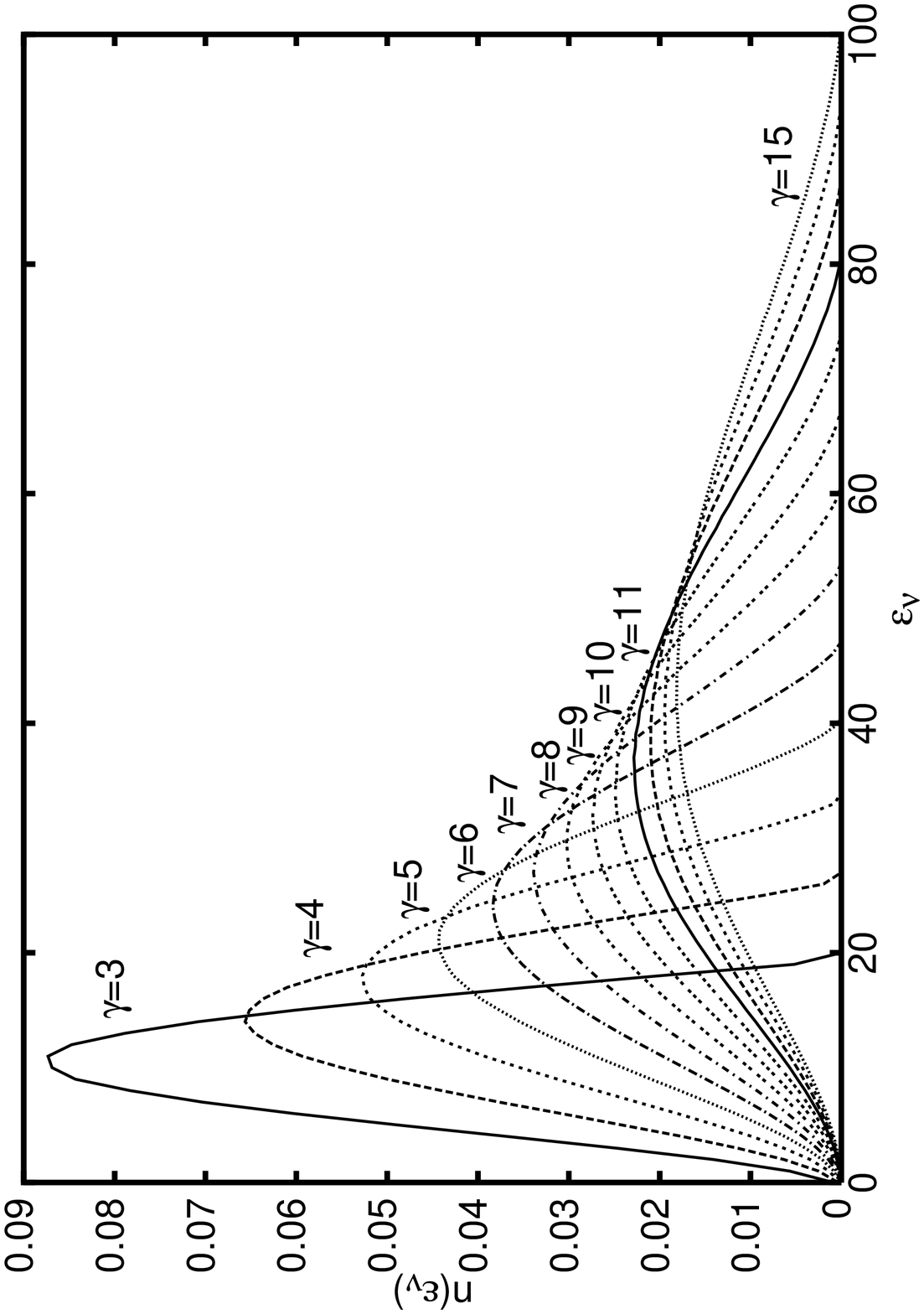"}
\caption{Normalized beta-beam neutrino-spectra stemming from $^{18}$Ne for different boost factors $\gamma$ between 3 and 15,  for a target with a cross sectional area of 4 m$^2$, located 10 m from one end of the ring, which has a straight side length of more than 90 m.} 
\label{spec3}
\end{figure} 

Energy distributions of supernova neutrinos are shaped by the circumstances in which the neutrinos are emitted.  Neutrinos leaving the star are responsible for the cooling of the proto-neutronstar forming in the star's core.
Hence, their spectrum resembles a thermal one, with temperatures reflecting the conditions at the site where they decoupled.  However, the fact that different kinds of neutrinos are involved in different interactions, and that  the reactivity of the (anti)neutrino depends on its energy, flavor and helicity, modulates this picture.  

For all (anti)neutrino flavors, the  energies
are in the range of a few to a few tens of MeV, although calculations
of neutrino transport which use different opacities achieve somewhat 
different spectra \cite{Liebendoerfer:2003es}.
Traditionally, Fermi-Dirac spectra were put forward as the convenient description for the energy distribution $n_{SN}(\varepsilon_{\nu_e})$ of supernova neutrinos
\begin{equation}
n_{FD[T,\eta]}(\varepsilon_{\nu})=\frac{N_{\eta}}{T^3}\frac{\varepsilon_{\nu}^2}{1+e^{\frac{\varepsilon_{\nu}}{T}-\eta}}.
\end{equation}
Recent results showed the supernova neutrino energy distribution to be accurately
parametrized with a power-law distribution \cite{janka1,janka2}
\begin{equation}
 n_{PL[\langle \varepsilon\rangle,\alpha]}(\varepsilon_{\nu})=\left(\frac{\varepsilon_{\nu}}{\langle \varepsilon_{\nu}\rangle}\right)^{\alpha}\,e^{-(\alpha+1)\frac{\varepsilon_{\nu}}{\langle \varepsilon_{\nu} \rangle}} ,
\end{equation}
where $T$ represents the temperature at the decoupling site of the neutrinos and  $\langle \varepsilon_{\nu}\rangle$ is the average neutrino energy. Up to zeroth order in $\eta$, both quantities are related by $T=3.15\: \langle \varepsilon_{\nu}\rangle$. In general, higher average energies or temperatures lead to broader spectra reaching a maximum at higher energy values, and with enhanced tails. The parameters $\eta$ and $\alpha$ allow to adjust the width  $w=\sqrt{\langle\varepsilon_{\nu}^2\rangle-\langle\varepsilon_{\nu}\rangle^2}$ of the spectrum.   Larger values for $\alpha$ or $\eta$ reduce the width of the spectrum and the influence of its tail. $N_{\eta}$ is a normalization factor depending on the width parameter.
The parameterizations allow to construct spectra that are equivalent up to their second moment by adjusting the energy and width parameters. 
These spectra are illustrated in Fig.~\ref{spec1}. 
The higher energies correspond to mu- and tau-neutrinos, interacting only through neutral-current interactions and decoupling close to the center of the star at relatively high temperatures. The low-energy spectra are important for electron neutrinos and antineutrinos, whose opacities are built from neutral- and charged-current interactions.
The precise shape of the spectrum and its tail is very important for the nuclear response to supernova neutrinos, as cross sections are rising fast with increasing neutrino energies. This is due to the fact that the supernova neutrino energy range is probing the giant resonance region of the nuclear spectrum, where cross sections are varying fast.  This makes nuclei very sensitive probes on one hand, but on the other hand makes the concept of nuclei as supernova-neutrino detectors very sensitive to uncertainties in nuclear structure calculations too.

The experimental neutrino beams mainly used up to now to study neutrino interactions at these energies, stem from pion decay at rest:
\begin{equation}
\pi^+\;\rightarrow\; \mu^+\;+\;\nu_{\mu},
\end{equation}
and the subsequent decay of the muon
\begin{equation}
\mu^+\;\rightarrow\; e^+\;+\;\nu_{e}\;+\;\overline{\nu}_{\mu}.
\end{equation}
The muon neutrinos have energies $\varepsilon_{\nu_{\mu}}=29.8$ MeV, the other neutrinos produced in these reactions have energies distributed according to the Michel spectrum~:
\begin{eqnarray}
n_{\nu_e}(\varepsilon_{\nu_e})&=&\frac{96\varepsilon_{\nu_e}^2}{m_{\mu}^4}(m_{\mu}-2\varepsilon_{\nu_e}),\\
n_{\overline{\nu}_{\mu}}(\varepsilon_{\overline{\nu}_{\mu}})&=&\frac{32\varepsilon_{\overline{\nu}_{\mu}}^2}{m_{\mu}^4}(\frac{3}{2}m_{\mu}-2\varepsilon_{\overline{\nu}_{\mu}}),
\end{eqnarray}
providing electron neutrinos and muon antineutrinos with energies up to 52.8 MeV. These spectra are illustrated in Fig.~\ref{spec2}.
Their energy covers the same range as that of supernova neutrinos, but 
the shape is rather different. Peak energies are much higher than those for supernova neutrinos and  the long tail, important for the study of supernova neutrinos, is lacking.
Fig.~\ref{spec3} shows that low-energy beta-beam neutrino energy distributions are covering the same energy range as that for supernova neutrinos, with a shape that is remarkably similar to that of supernova-neutrino spectra.

\section{Reconstructing neutrino spectra and nuclear responses}
\begin{figure*}[htb]
\vspace*{12.5cm}
\special{hscale=35 vscale=35 hsize=1500 vsize=600
         hoffset=-15 voffset=380 angle=-90 psfile="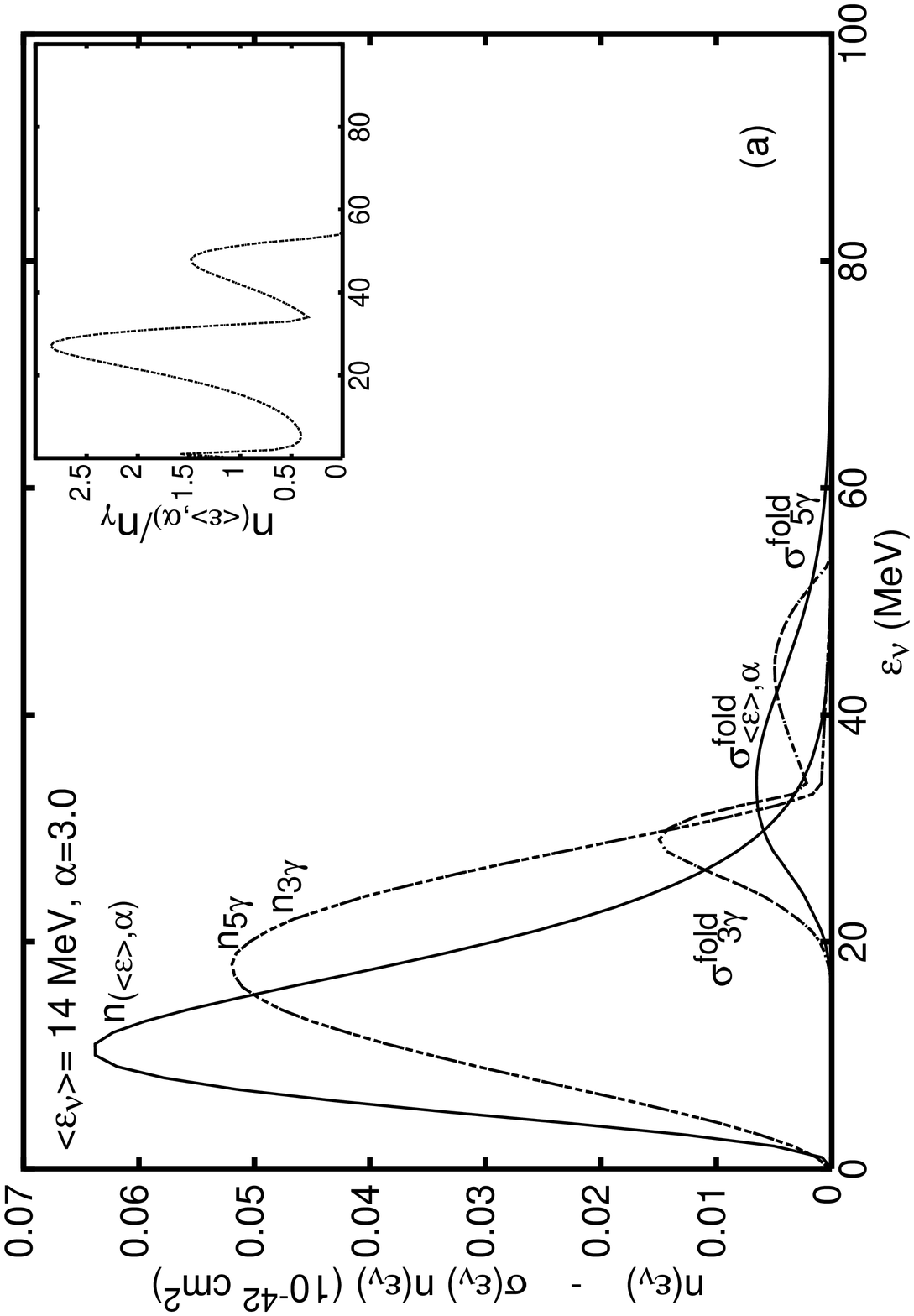"}
\special{hscale=35 vscale=35 hsize=1500 vsize=600
         hoffset=235 voffset=380 angle=-90 psfile="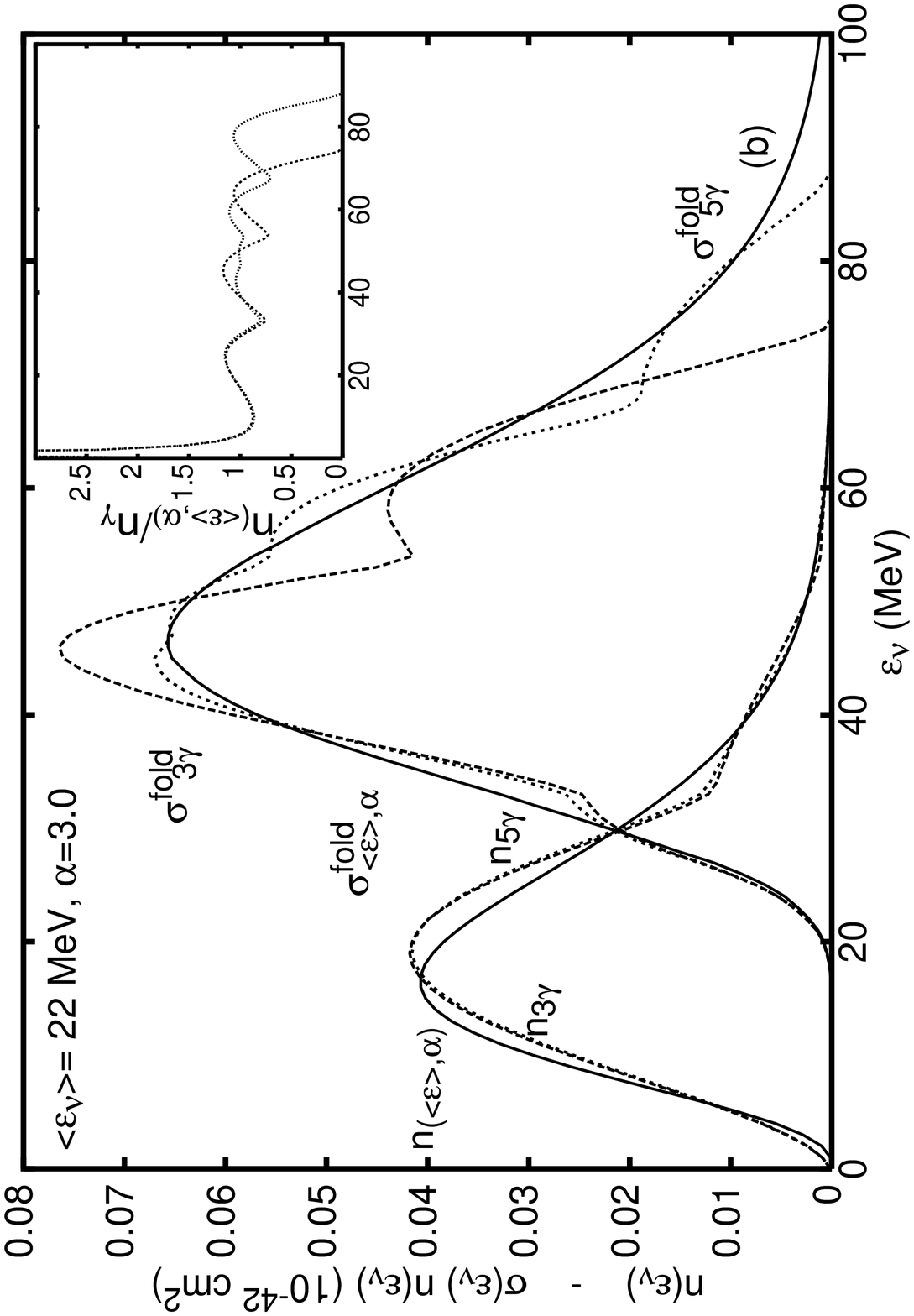"}
\special{hscale=35 vscale=35 hsize=1500 vsize=600
         hoffset=-15 voffset=200 angle=-90 psfile="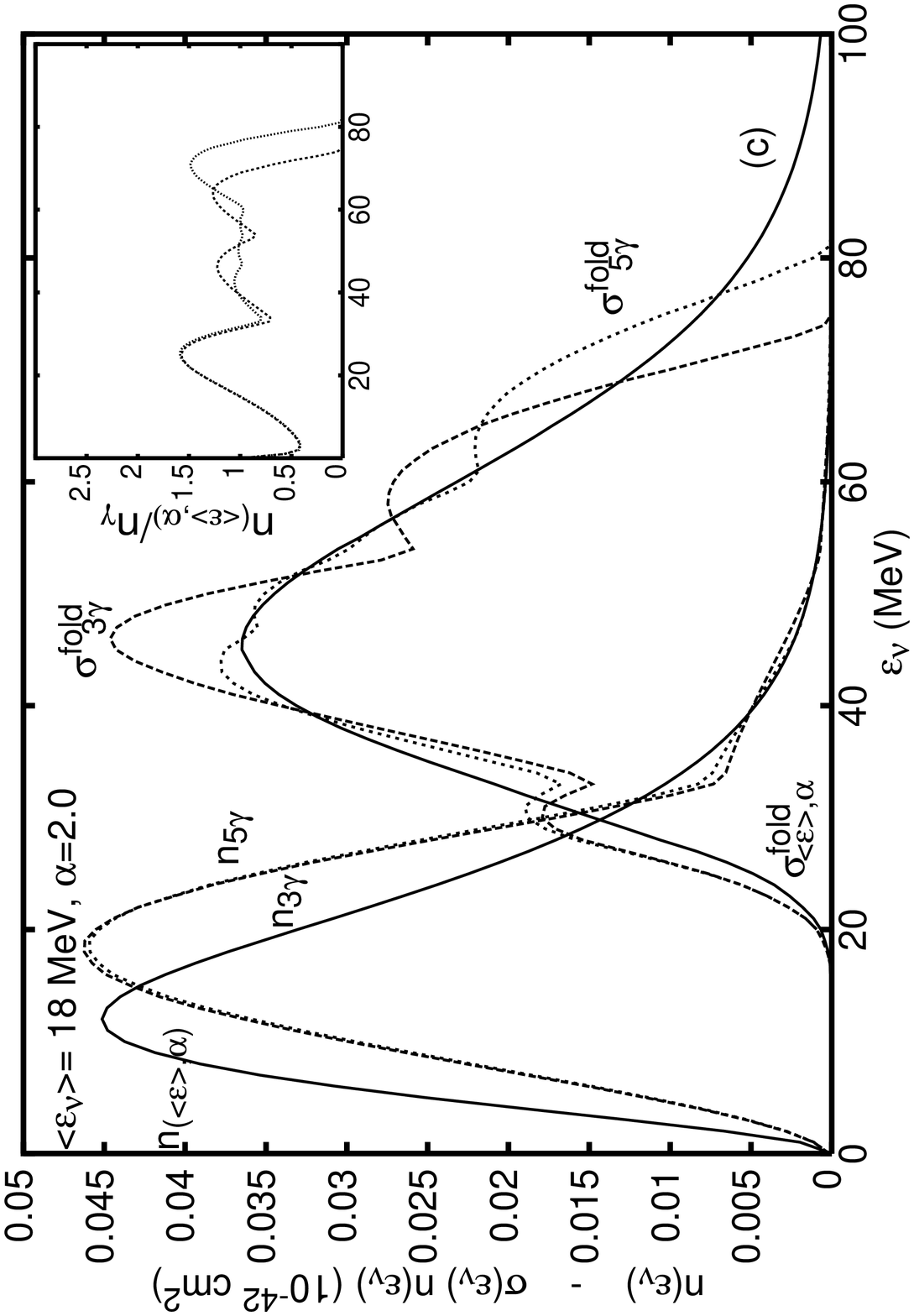"}
\special{hscale=35 vscale=35 hsize=1500 vsize=600
         hoffset=235 voffset=200 angle=-90 psfile="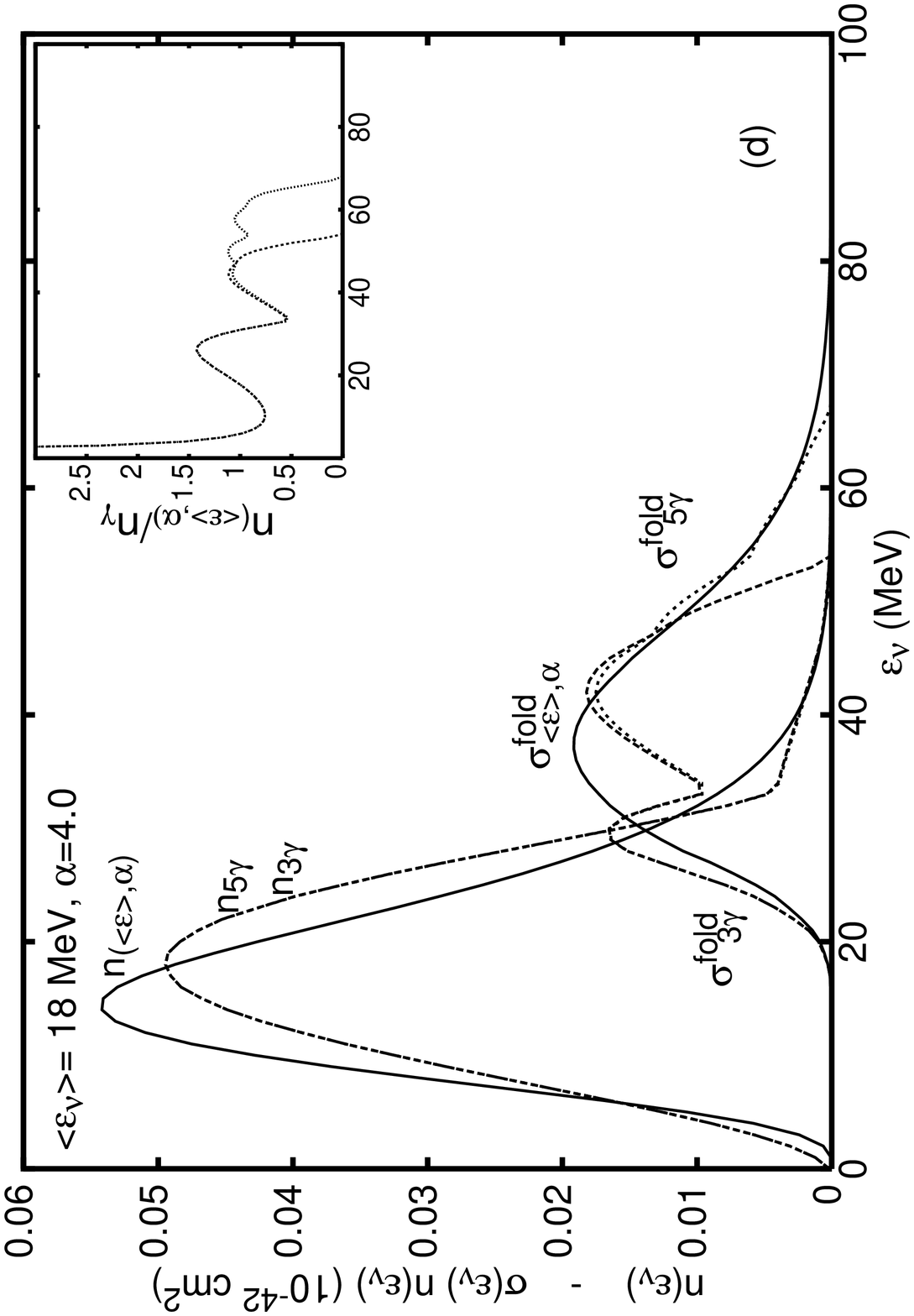"}
\caption{In each plot, the curves on the left show the original
  supernova-neutrino spectrum $n_{(<\varepsilon>, \alpha)}$ (full
  line), and the fit with three beta-beam spectra at different gamma
  $n_{3\gamma}$ (dashed), and 5 beta-beam spectra $n_{5\gamma}$
  (dotted).  The curves on the right show the  $^{16}O(\nu,\nu')
  ^{16}O^*$ cross section, folded with the SN neutrino spectrum
  $\sigma^{fold}_{<\varepsilon>, \alpha}$  (full line),  the
  artificial spectrum constructed using 3 different beta-beam spectra
  $\sigma^{fold}_{3\gamma}$ (dashed), and the artificial spectrum
  built from 5 beta-beam spectra at  different gammas $\sigma^{fold}_{5\gamma}$ (dotted). The cross section was obtained with a continuum random phase approximation calculation \cite{mepb}.  The inset shows the ratio of constructed spectrum to the original one for the synthetic spectrum containing 3  (dashed line) and 5 beta-beam spectra (dotted).}
\label{fit}
\end{figure*}

We aim at  optimizing the information that low-energy beta-beams  can provide  about supernova-neutrino interactions and the interpretation of a supernova-neutrino signal in a terrestrial detector
To this goal we construct normalized linear combinations $n_{N\gamma}$ of beta-beam spectra $n_{\gamma_i}$ :
\begin{equation}
n_{N\gamma}(\varepsilon_{\nu})=\sum_{i=1}^N a_i n_{\gamma_i}(\varepsilon_{\nu}),
\end{equation}
with 
\begin{equation}
\int d\varepsilon_{\nu}\: n_{N\gamma}(\varepsilon_{\nu})=1,
\end{equation}
and
\begin{equation}
\int d\varepsilon_{\nu}\: n_{\gamma_i}(\varepsilon_{\nu})=1 \,\,\,\,\,\,\, \forall i,
\end{equation} 
and varied the boost factors $\gamma_{i=1,\cdots,N}$  and the expansion coefficients $a_{i=1,\cdots,N}$ to minimize the expression
\begin{equation}
\int_{\varepsilon_{\nu}} d\varepsilon_{\nu} \left|n_{N\gamma}(\varepsilon_{\nu})-n_{SN}(\varepsilon_{\nu})\right|. \label{minim}
\end{equation}
In this way we obtain a synthetic spectrum $n^{fit}_{N\gamma}(\varepsilon_{\nu})$ that is the best fit to the supernova-neutrino energy-distribution $n_{SN}(\varepsilon_{\nu})$ for particular values of the average energy and width.
Using the norm of the expression $\left|n_{N\gamma}(\varepsilon_{\nu})-n_{SN}(\varepsilon_{\nu})\right|$, rather than the square of the norm, we avoid  giving  a larger weight to the peak of the distribution. This is important in view of the fact that the spectrum's tail is more important to our applications.

Fig.~\ref{fit} shows some results of best fits to power-law spectra
using N=3 and N=5 beta-beam spectra in the synthetic energy distribution,
for different supernova-neutrino energy-distributions. The $\gamma$'s
were allowed to assume  integer values between 5 and 15.  The
resulting synthetic spectra are  compared with the original supernova
spectrum, and the corresponding folded cross sections
\begin{equation}
\sigma^{fold}_{N\gamma}(\varepsilon_{\nu})= \sigma(\varepsilon_{\nu})n_{N\gamma}(\varepsilon_{\nu}),
\end{equation}
and
\begin{equation}
\sigma^{fold}_{<\varepsilon>,\alpha}(\varepsilon_{\nu})= \sigma(\varepsilon_{\nu})n_{(<\varepsilon>,\alpha)}(\varepsilon_{\nu}),
\end{equation}
are confronted.  These are especially important 
because the energy dependence of the yield of neutrino interactions scales directly
with the folded cross section.

The overall fit is quite good.  It is clear that, especially for spectra with lower average energy, the synthetic spectra tend to peak at slightly higher energies.  This is due to the fact that the beta-beam spectrum at $\gamma$=5, the lowest one included in the fit, is just slightly higher in energy than the lowest predictions for supernova neutrino energy distributions are.   The procedure does not use the freedom to include higher gammas in the constructed spectrum, the fit merely seeks for better agreement at low energies.

It is   important to emphasize that the fitting procedure
is very powerful in reproducing the high-energy tail of the supernova neutrino energy spectrum.  The inset in Fig.~\ref{fit} illustrates that, whereas for incoming neutrino energies below and around 20 MeV, the synthetic spectra are still wobbling around the original one, for energies above 30 MeV the fit becomes much more stable. Especially for constructed spectra including 5 gamma distributions, the agreement is very good.
The fact that the synthetic spectrum peaks at slightly higher energies than the original one  does  not limit the strength of the proposed procedure.  The important quantity to reproduce is not the spectrum, but the cross section, folded with the spectrum.  
This folded cross section determines the nuclear response to supernova neutrinos.  It reaches a maximum around 30 to 40 MeV, indicating that neutrinos with these energies 
have the largest impact on the nuclear response \cite{Jachowicz:2003iz}.   
The bunch of curves right in 
Fig.~\ref{fit} shows cross sections as a function of the incoming neutrino energy, folded with the supernova neutrino spectra, and folded with the constructed spectra. The folded cross-section curves  pop up only in the very tail of the energy distribution, and are in  very nice agreement.

\begin{figure*}[htb]
\vspace*{6.5cm}
\special{hscale=35 vscale=35 hsize=1500 vsize=600
         hoffset=-10 voffset=200 angle=-90 psfile="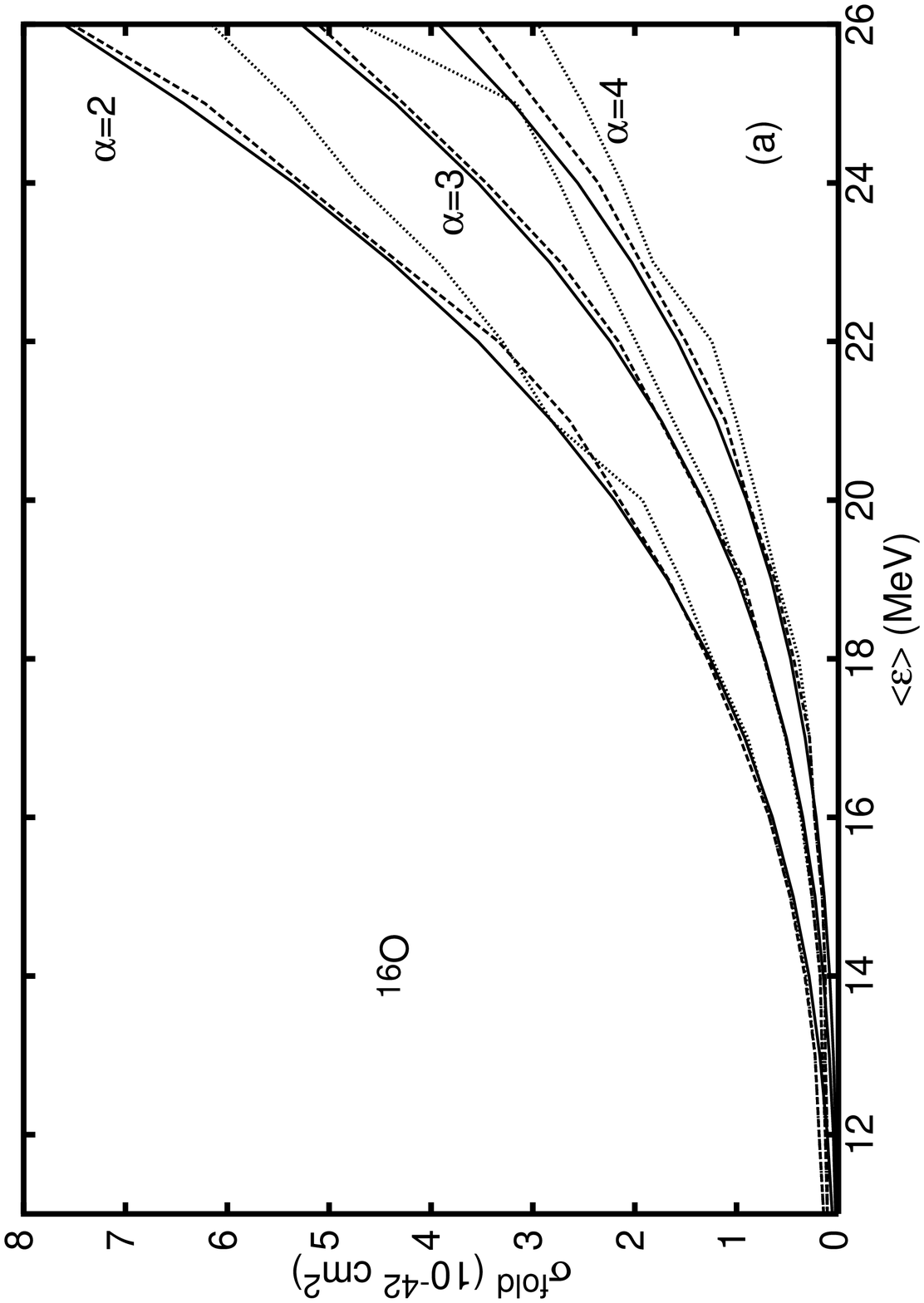"}
\special{hscale=35 vscale=35 hsize=1500 vsize=600
         hoffset=235 voffset=200 angle=-90 psfile="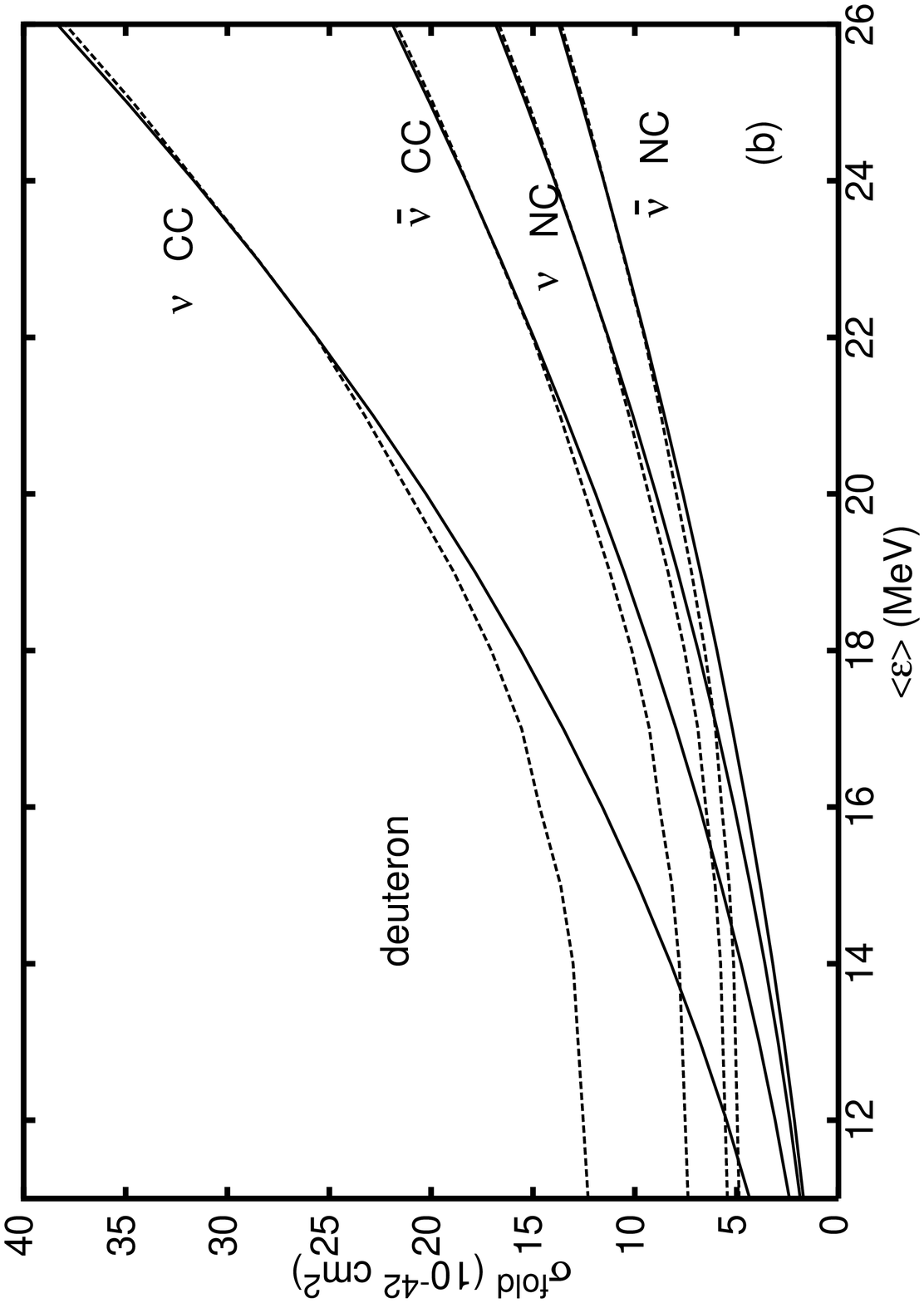"}
\caption{Total cross sections for neutrino scattering on $^{16}$O (a), and on the deuteron (b), folded with different spectra. The $^{16}$O results were taken from Ref.~\cite{menc}.  The folded cross sections are shown for the supernova neutrino spectrum (full line), a synthetic spectrum with 3 (dotted line) and with 5 gamma spectra (dashed lines) in the construction. The different bunches of curves correspond to different widths of the energy distributions, as indicated by the parameter $\alpha$.  For the deuteron, cross sections were taken from Ref.~\cite{kubo}.
The cross sections folded with the original spectra (full line) and with the fit consisting of 5 gamma spectra (dashed line) are shown for neutral and charged-current neutrino and antineutrino scattering.  The width of the energy spectrum was kept fixed at $\alpha=3$ in this case.} 
\label{tot}
\end{figure*}

Fig.~\ref{tot} compares total folded cross sections for supernova spectra $\sigma_{SN}=\int d\varepsilon_{\nu}\: \sigma(\varepsilon_{\nu}) n_{SN}(\varepsilon_{\nu})$ and the equivalent synthetic folded cross sections
\begin{eqnarray}
\sigma^{fit}_{N\gamma}&=&\int d\varepsilon_{\nu}\: \sigma(\varepsilon_{\nu}) n^{fit}_{N\gamma}(\varepsilon_{\nu})\\
&=&\sum_{i=1}^N a^{fit}_i \int d\varepsilon_{\nu}\: \sigma(\varepsilon_{\nu})n_{\gamma_i}^{fit}(\varepsilon_{\nu}).
\end{eqnarray}
For $^{16}$O, the overall agreement is very good, supporting the strength of the proposed procedure.  For the deuteron, the cross sections  are much more sensitive to lower energies than those for massive nuclei are, and for spectra with small average energies the fit overestimates the folded cross sections.  The agreement becomes better between 16 and 18 MeV average energy, growing to an almost perfect match for the highest energies we examined.

The fact that the technique is already very efficient in the straightforward  way it was presented, leaves opportunity for optimization. In the following paragraphs, we suggest a number of ways to make the technique more efficient. 

\begin{figure*}[htb]
\vspace*{6.5cm}
\special{hscale=35 vscale=35 hsize=1500 vsize=600
         hoffset=-10 voffset=200 angle=-90 psfile="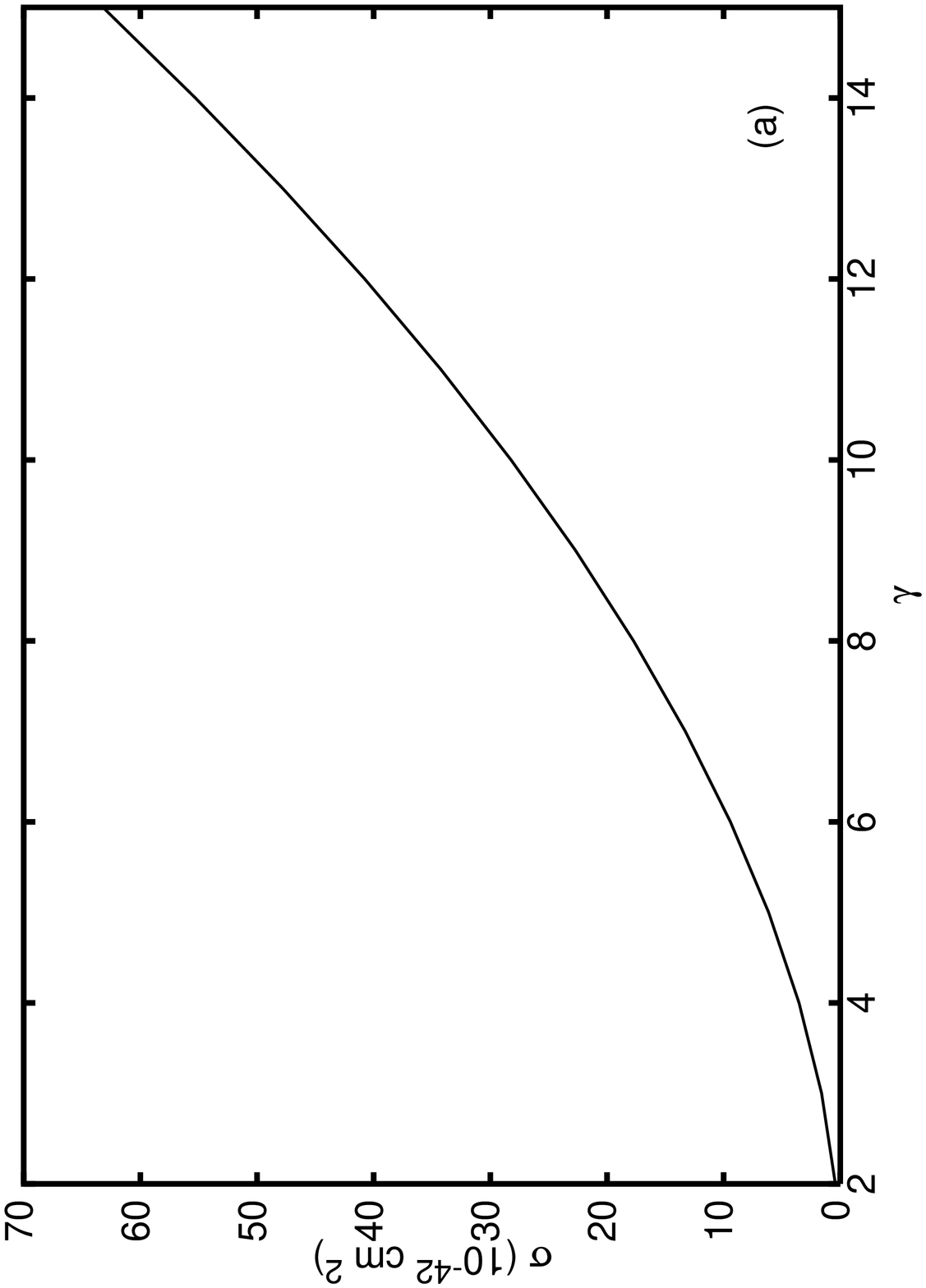"}
\special{hscale=35 vscale=35 hsize=1500 vsize=600
         hoffset=235 voffset=200 angle=-90 psfile="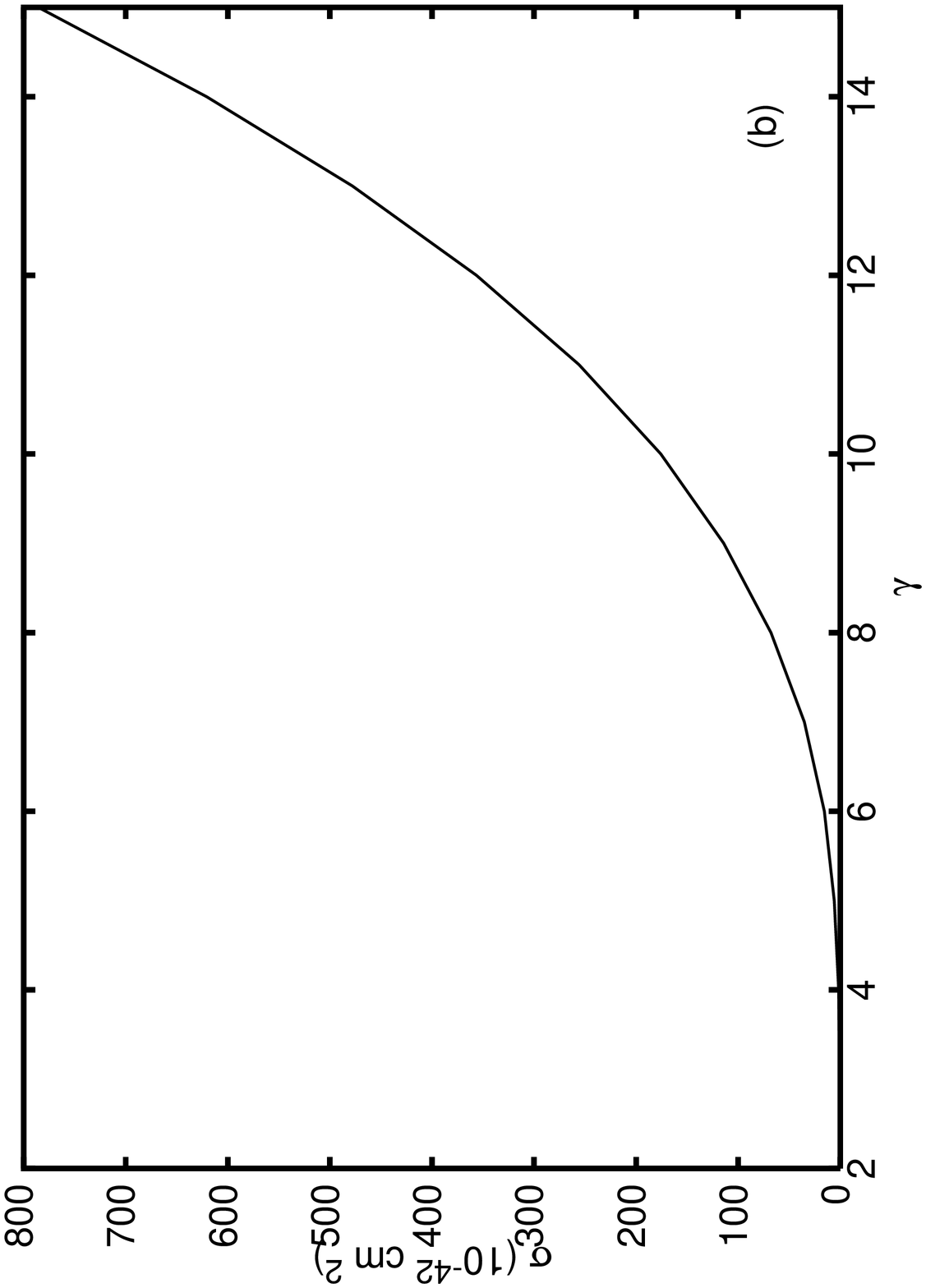"}
\caption{Folded cross section as a function of $\gamma$. The left panel shows the results for neutral current neutrino scattering on the deuteron $\nu$d$\rightarrow$$\nu'$pn, the right panel for neutral current scattering on  $^{208}$Pb.} 
\label{gamfig}
\end{figure*}

(1) Figure \ref{gamfig} shows that the folded cross sections behave very smoothly as a function of the boost factor $\gamma$.  This is a major advantage for the proposed procedure~: a number of measurements at particular gammas will allow to infer other values.  Thus, the fitting procedure can be offered more freedom in its choice of $\gamma$, 
 which will definitely result in an even better agreement.  Moreover the curves in Fig.~\ref{gamfig} might be used as guide to extrapolate cross sections to lower gamma values, where the event rate in the detector becomes small.  All low-energy fits, and especially the synthetic spectra  for the deuteron, would benefit from the inclusion of gammas smaller than 5  in the fit.

\begin{figure}[htb]
\vspace*{7.cm}
\special{hscale=35 vscale=35 hsize=1500 vsize=600
         hoffset=-10 voffset=200 angle=-90 psfile="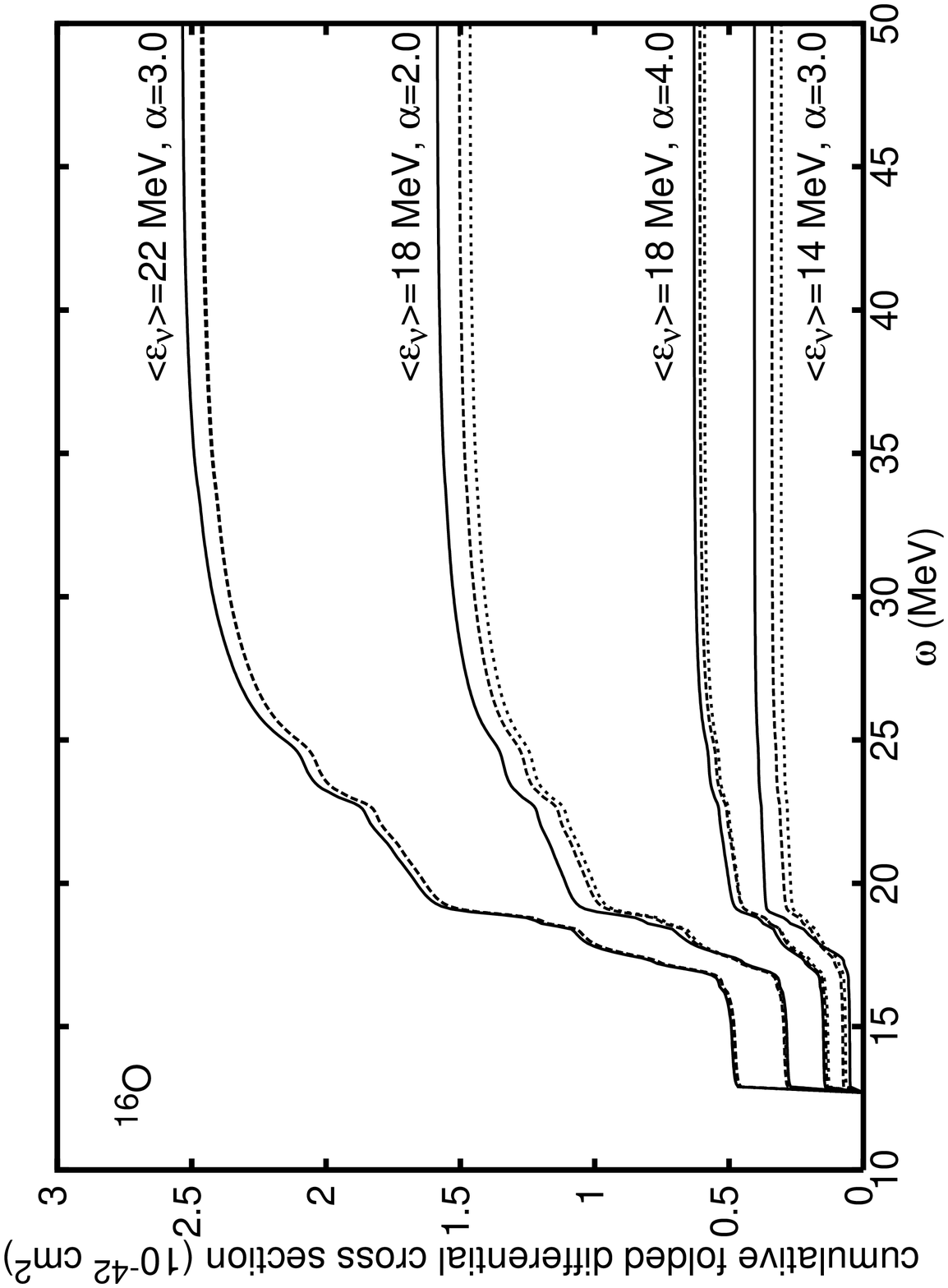"}
\caption{Folded cumulative cross section for neutral current neutrino scattering off $^{16}$O as a function of the excitation energy $\omega$. The figure compares cross sections folded with different spectra : original power law spectra (full line), the synthetic spectrum obtained without weight in the fit (dotted), and the fit with the cross section calculation used as weight (dashed line).} \label{csfig}
\end{figure}

(2) Once one has experimental beta-beam data available, the fitting procedure of Eq.~\ref{minim} can be optimized by including  an energy-dependent weight function in the minimization process :
\begin{equation}
\int_{\varepsilon_{\nu}} d\varepsilon_{\nu} \left|n_{N\gamma}(\varepsilon_{\nu})-n_{SN\gamma}(\varepsilon_{\nu})\right|w(\varepsilon_{\nu}).
\end{equation}
In this way, it becomes possible to ensure that the obtained agreement is best in important energy regions i.e. the match between original spectrum and fit should be closest in energy regions where the cross section is high. As a matter of fact, once experimental values of the folded cross section are available, the optimum choice for the weight would be a cross section calculation $\sigma^{th}$, translating the minimization procedure in 
\begin{equation}
\int_{\varepsilon_{\nu}} d\varepsilon_{\nu} \left|n_{N\gamma}(\varepsilon_{\nu})-n_{SN\gamma}(\varepsilon_{\nu})\right|\sigma^{th}(\varepsilon_{\nu}),
\end{equation}
or
\begin{equation}
\int_{\varepsilon_{\nu}} d\varepsilon_{\nu} \left|\sigma^{fold}_{N\gamma}(\varepsilon_{\nu})-\sigma^{fold}_{SN\gamma}(\varepsilon_{\nu})\right|.
\end{equation}
The synthetic spectrum $n_{N\gamma}^{fit}(\varepsilon_i)$ obtained with this procedure  serves as a guide for the choice of gammas in the experiment. The appropriate combination of the responses in the detector then provides a very good reproduction of the supernova-neutrino signal. This is illustrated in Fig.~\ref{csfig}.
Especially for spectra with smaller energies, using an energy-weighted fit improves the results of the procedure. 

\begin{table*}
\begin{center}
\begin{tabular}{cc||ccccccccccc|c}
\hline
$\langle E\rangle$&$\alpha$&$a_{\gamma=5}$&$a_{\gamma=6}$&$a_{\gamma=7}$&$a_{\gamma=8}$&$a_{\gamma=9}$&$a_{\gamma=10}$&$a_{\gamma=11}$&$a_{\gamma=12}$&$a_{\gamma=13}$&$a_{\gamma=14}$&$a_{\gamma=15}$& $\int_{\varepsilon_{\nu}} \!\!\!\!d\varepsilon_{\nu}|n^{fit}(\varepsilon_{\nu})\!-\!n_{SN}(\varepsilon_{\nu})|$\\\hline
14&2&0.9355& 0.0000& 0.0003& 0.0502& 0.0074& 0.0043& 0.0013& 0.0008& 0.0000& 0.0001& 0.0000& 0.5564
 
\\

14&3&0.9745& 0.0000& 0.0000& 0.0216& 0.0027& 0.0009& 0.0002& 0.0000& 0.0000& 0.0000& 0.0000&0.5306

\\
14&4& 0.9904& 0.0000& 0.0000& 0.0084& 0.0011& 0.0000& 0.0000& 0.0000& 0.0000& 0.0000& 0.0000&0.5293
\\
18&2&0.7306& 0.0001& 0.1180& 0.0671& 0.0375& 0.0244& 0.0089& 0.0088& 0.0015& 0.0008& 0.0022&0.3078

\\
18&3&0.8221& 0.0000& 0.0626& 0.0768& 0.0179& 0.0137& 0.0035& 0.0029& 0.0000& 0.0002& 0.0003&0.2352
\\
18&4&0.8577& 0.0000& 0.0741& 0.0501& 0.0097& 0.0067& 0.0012& 0.0002& 0.0002& 0.0001& 0.0000&0.2048
 
\\
22&2& 0.5468& 0.0001& 0.1359& 0.1075& 0.0671& 0.0599& 0.0273& 0.0273& 0.0084& 0.0004& 0.0192&0.1695

\\
22&3&0.5563& 0.0000& 0.1762& 0.1097& 0.0726& 0.0380& 0.0272& 0.0074& 0.0082& 0.0008& 0.0035&0.0915

\\
22&4&0.4969& 0.0783& 0.2184& 0.0931& 0.0663& 0.0213& 0.0187& 0.0042& 0.0013& 0.0011& 0.0005&0.0803

\\
\hline
\end{tabular}
\caption{Expansion parameters $a_{\gamma=5,...,15}$ for the fit to the power-law supernova-neutrino spectrum, defined by average energy $\langle \varepsilon \rangle$ and width $\alpha$. The quantity $\int_{\varepsilon_{\nu}} d\varepsilon_{\nu}|n^{fit}(\varepsilon_{\nu})-n_{SN}(\varepsilon_{\nu})|$ in the last column provides a measure for the goodness of fit. 
\label{table111}}
\end{center}
\end{table*}

\begin{table*}
\begin{center}
\begin{tabular}{cc||cccccccccc|c}
\hline
$\langle E\rangle$&$\alpha$
&$\begin{array}{c}\gamma=7\\(2.13,5)\end{array}\!\!\!\!\!\!$
&$\begin{array}{c}\gamma=7\\(2.75,3)\end{array}\!\!\!\!\!\!$
&$\begin{array}{c}\gamma=7\\(3.89,1.5)\end{array}\!\!\!\!\!\!$
&$\begin{array}{c}\gamma=7\\(4.77,1)\end{array}\!\!\!\!\!\!$
&$\begin{array}{c}\gamma=7\\(6.74,0.5)\end{array}\!\!\!\!\!\!$
&$\begin{array}{c}\gamma=14\\(2.13,5)\end{array}\!\!\!\!\!\!$
&$\begin{array}{c}\gamma=14\\(2.75,3)\end{array}\!\!\!\!\!\!$
&$\begin{array}{c}\gamma=14\\(3.89,1.5)\end{array}\!\!\!\!\!\!$
&$\begin{array}{c}\gamma=14\\(4.77,1)\end{array}\!\!\!\!\!\!$
&$\begin{array}{c}\gamma=14\\(6.74,0.5)\end{array}\!\!\!$
&$\int_{\varepsilon_{\nu}} \!\!\!\!d\varepsilon_{\nu}|n^{fit}(\varepsilon_{\nu})\!-\!n_{SN}(\varepsilon_{\nu})|$\\\hline
14&2&0.0000& 0.0000& 0.0000& 0.0000& 0.9982& 0.0000& 0.0000& 0.0001& 0.0003& 0.0014&0.5538 
\\
14&3&0.0000& 0.0000& 0.0000& 0.0000& 1.0000& 0.0000& 0.0000& 0.0000& 0.0000& 0.0000&0.5794

\\
14&4&  0.0000& 0.0000& 0.0000& 0.0000& 1.0000& 0.0000& 0.0000& 0.0000& 0.0000& 0.0000&
0.6274

\\
18&2&0.0000& 0.0000& 0.0000& 0.0000& 0.9764& 0.0000& 0.0000& 0.0002& 0.0014& 0.0219&
0.2595

\\
18&3&0.0000& 0.0000& 0.0000& 0.0000& 0.9970& 0.0001& 0.0009& 0.0020& 0.0000& 0.0000&0.2724
\\
18&4&0.0000& 0.0000& 0.0000& 0.0000& 0.9995& 0.0002& 0.0002& 0.0000& 0.0000& 0.0000&
 0.3276

\\
22&2& 0.0000& 0.0000& 0.0000& 0.0000& 0.8343& 0.0000& 0.0000& 0.0000& 0.0000& 0.1656&
0.1083
\\
22&3&0.0000& 0.0000& 0.4407& 0.0001& 0.4861& 0.0000& 0.0000& 0.0000& 0.0001& 0.0729&
0.1352

\\
22&4&0.0000& 0.0000& 0.9807& 0.0000& 0.0000& 0.0001& 0.0007& 0.0184& 0.0001& 0.0000&
0.1753
\\
\hline
\end{tabular}
\caption{Expansion parameters for the fit to  supernova-neutrino spectra, using the fluxes from Ref.~\cite{Amanik:2007zy}, obtained at different parts of a detector placed 10 m away from a 450 m storage ring with a straight section of 150 m.
All considered parts of the detector are cylindrical and  have a volume of 71.2 m$^3$, radius and thickness of the  parts are given beteween brackets.  The quantity $\int_{\varepsilon_{\nu}} d\varepsilon_{\nu}|n^{fit}(\varepsilon_{\nu})-n_{SN}(\varepsilon_{\nu})|$ in the last column provides a measure for the goodness of fit.\label{table112} }
\end{center}
\end{table*}

\begin{figure*}[hbt]
\vspace*{12.5cm}
\special{hscale=35 vscale=35 hsize=1500 vsize=600
         hoffset=-15 voffset=380 angle=-90 psfile="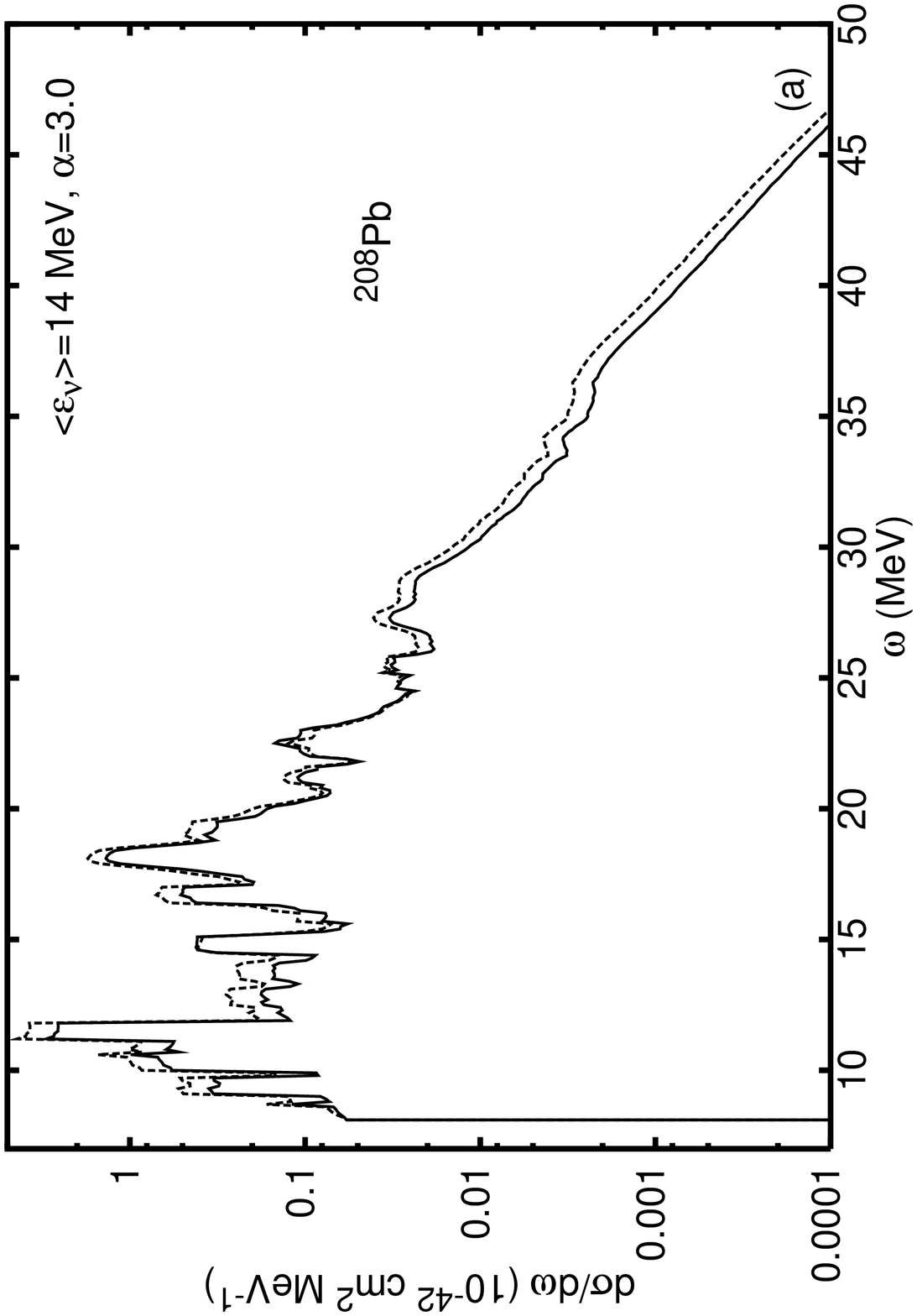"}
\special{hscale=35 vscale=35 hsize=1500 vsize=600
         hoffset=235 voffset=380 angle=-90 psfile="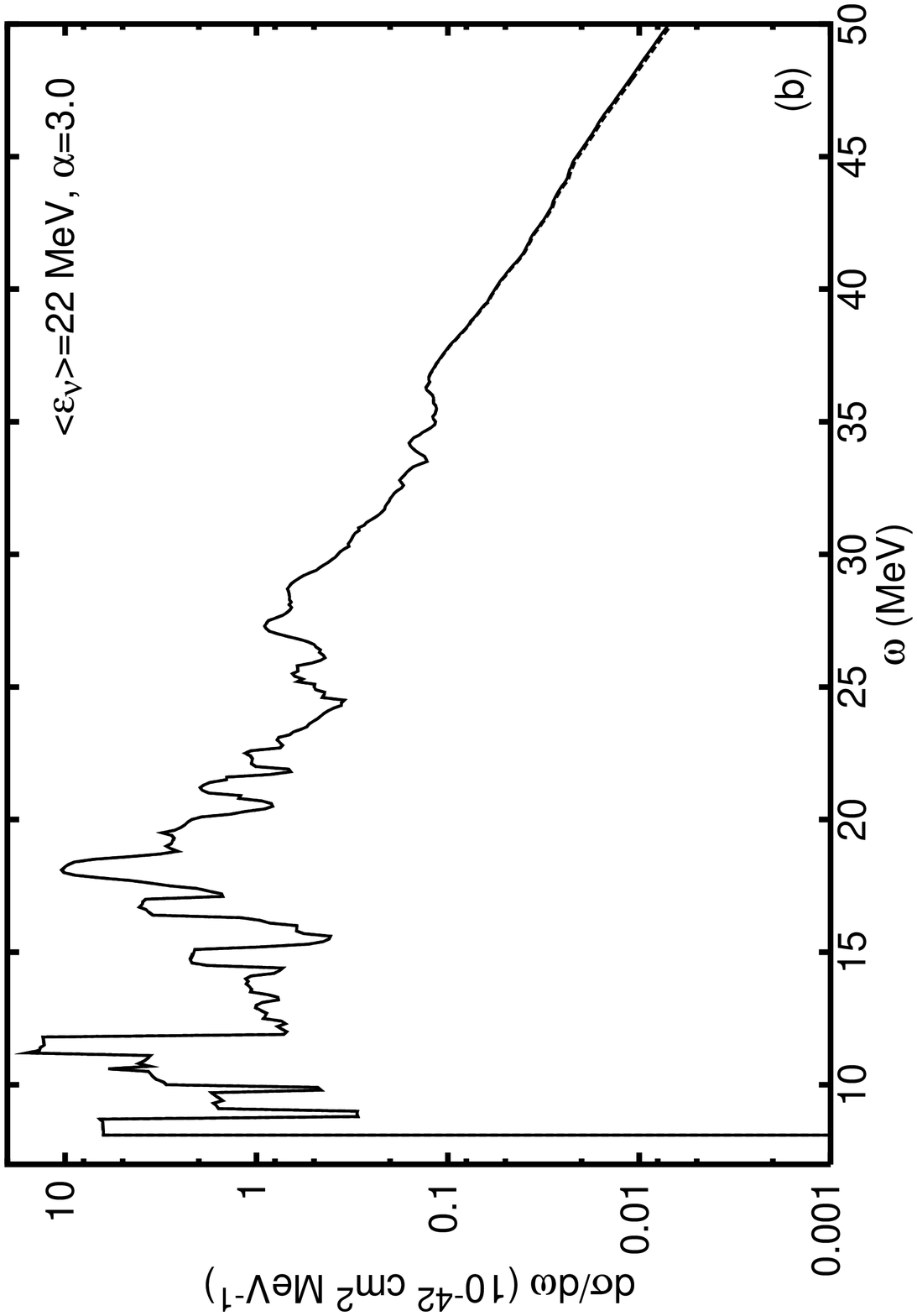"}
\special{hscale=35 vscale=35 hsize=1500 vsize=600
         hoffset=-15 voffset=200 angle=-90 psfile="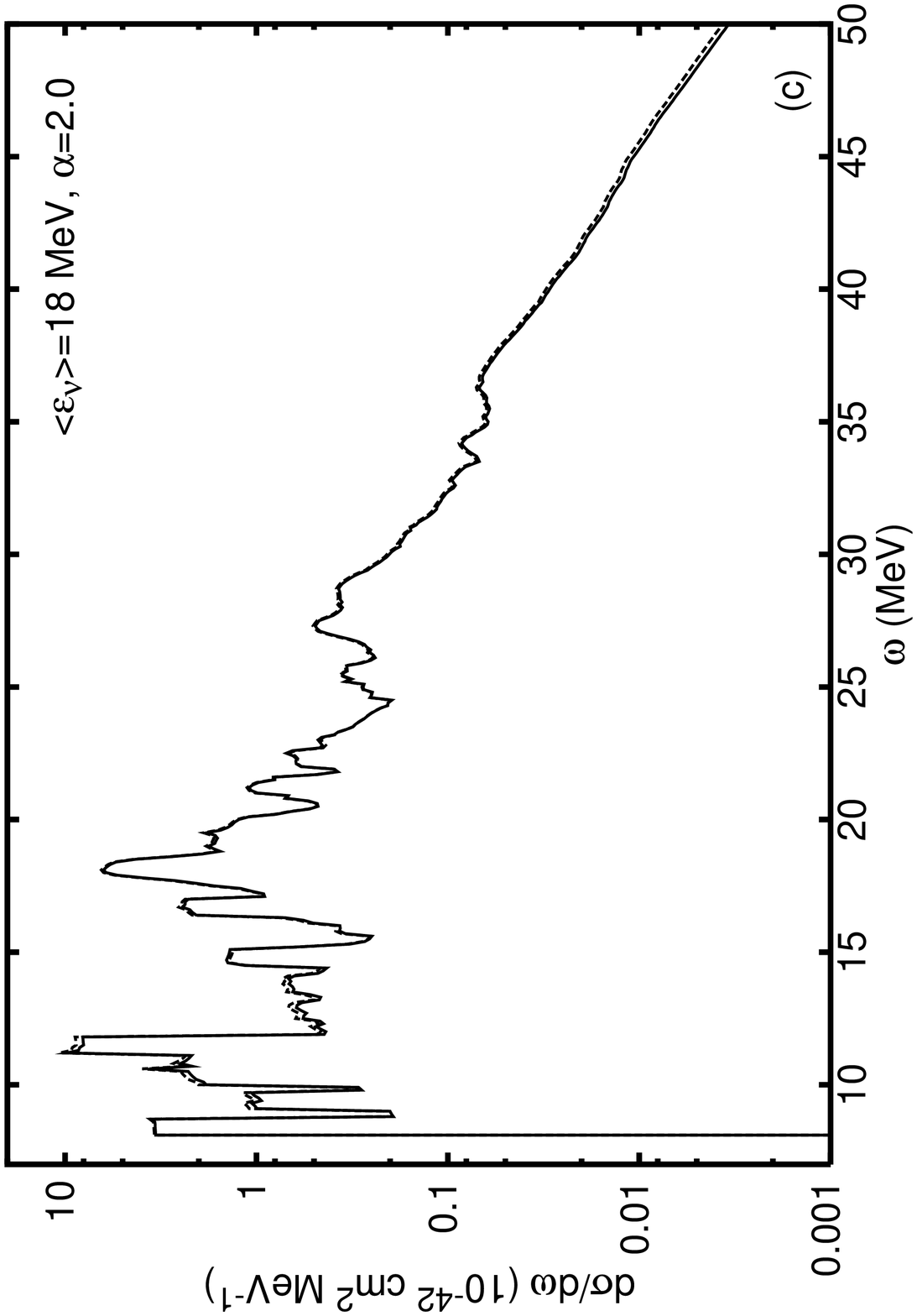"}
\special{hscale=35 vscale=35 hsize=1500 vsize=600
         hoffset=235 voffset=200 angle=-90 psfile="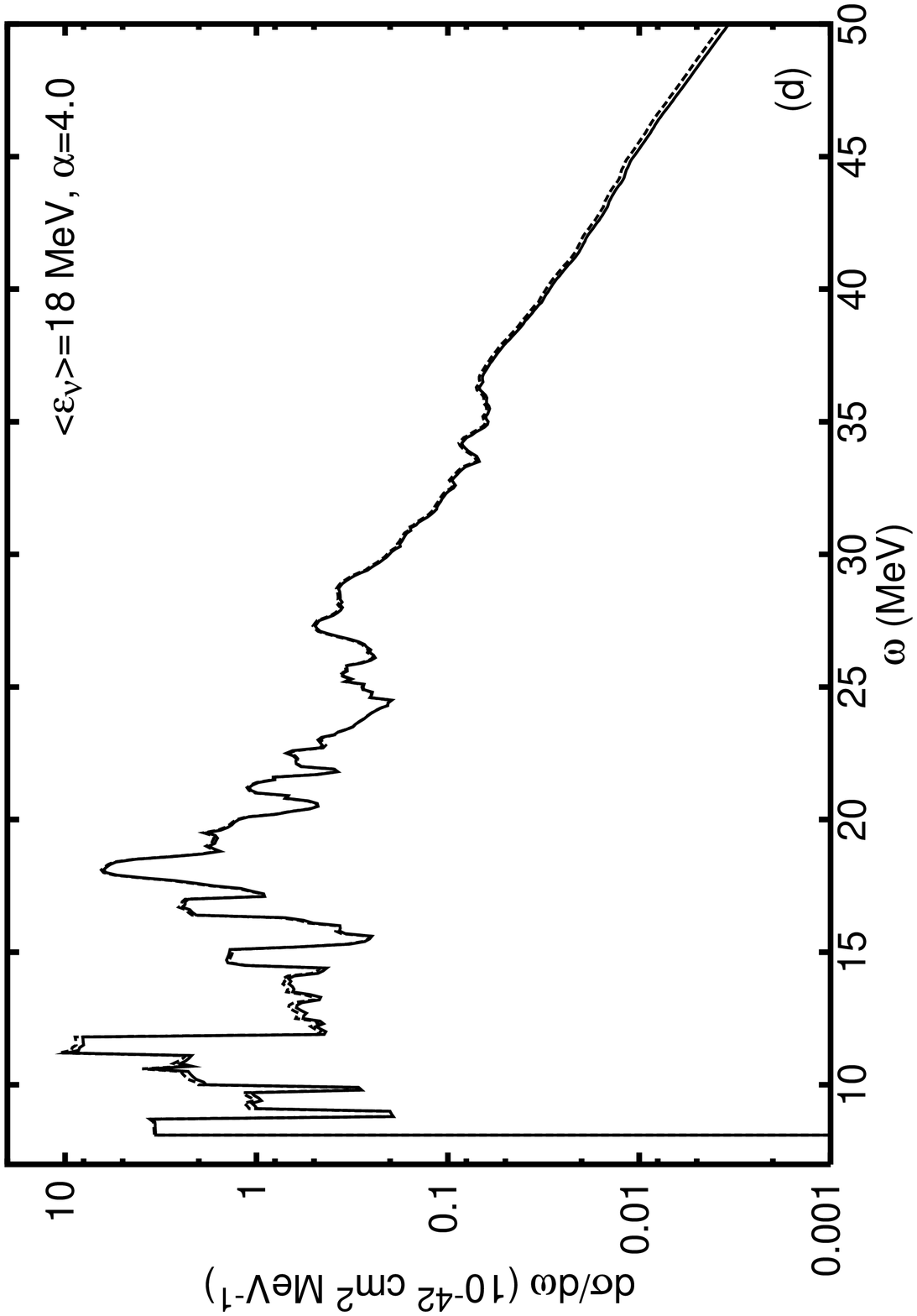"}
\caption{Comparison between differential cross sections for neutral current scattering on $^{208}$Pb, folded with a power-law supernova-neutrino spectrum (full line) and synthetic spectra taken from Tab.~\ref{table111} (dashed line)  with $\beta$-beam components  between $\gamma$=5 and 15,  for different energy distributions : $\langle\varepsilon\rangle$=14 MeV, $\alpha$=3 (a), $\langle\varepsilon\rangle$=22 MeV, $\alpha$=3 (b), $\langle\varepsilon\rangle$=18 MeV, $\alpha$=2 (c), and $\langle\varepsilon\rangle$=18 MeV, $\alpha$=4 (d). Cross sections were taken from \cite{mepb}.}
\label{fitpbfig}
\end{figure*}

(3) Up to now, we have shown that these responses can be reproduced using only 3 or 5 $\beta$-beam measurements at different $\gamma$. This restriction is rather academic.  The smooth behavior of the response as a function of the boost factor $\gamma$ assures that interpolating experimental results would not introduce extra uncertainties. Hence the proposed procedure will mainly benefit from the advantages   using a larger set of expansion functions brings along.
Table~\ref{table111} shows some results of best fits to power-law spectra, for different supernova-neutrino energy-distributions, where all $\gamma$ values between 5 and 15 were allowed to contribute to the synthetic spectra. As mentioned above, the most important quantity to study is the differential folded cross section i.e.~the folded cross section as a function of the final energy of the target nucleus.  This provides a measure for the energy transfer from the neutrinos to the material they are interacting with, and it determines the reaction products in nucleosynthesis processes. For supernova neutrino detection, the differential folded cross sections  indicate what one will 'see' in the detector, as the energy transfer and the excitation energy of the nucleus determine the decay products that will be observed in the detector. Fig.~\ref{fitpbfig} illustrates these response in terms of differential cross sections for neutral-current scattering off $^{208}$Pb, showing an almost perfect agreement between original and fitted response where all $\gamma$-values between 5 and 15 contribute to the synthetic spectra. As a matter of fact, the figure predicts the supernova-neutrino signal in a terrestrial detector and the response  on nuclei of interest in the  neutrino nucleosynthesis processes in terms of beta-beam responses.

(4) Table~\ref{table111} shows that the fit to the lowest-energy supernova neutrino spectra is dominated by beta-beam spectra at very low gammas.  From an experimental point of view, it might be beneficial  to use fluxes obtained at an off-axis detector or at different parts of the detector as an effective way to lower the average energy of the neutrinos, rather than running a facility at very low boost factors.  Table~\ref{table112} uses the fluxes from Ref.~\cite{Amanik:2007zy} obtained at different parts of a detector to vary the average energy and shape of the synthetic spectra, and shows that this is an excellent way to obtain a fit with the same accuracy level for a beta-beam facility run at higher gammas.

\section {Terrestrial detection of supernova neutrinos and neutrino oscillations}\label{oscsec}

\begin{figure}[htb]
\vspace*{6.5cm}
\special{hscale=35 vscale=35 hsize=1500 vsize=600
         hoffset=-10 voffset=200 angle=-90 psfile="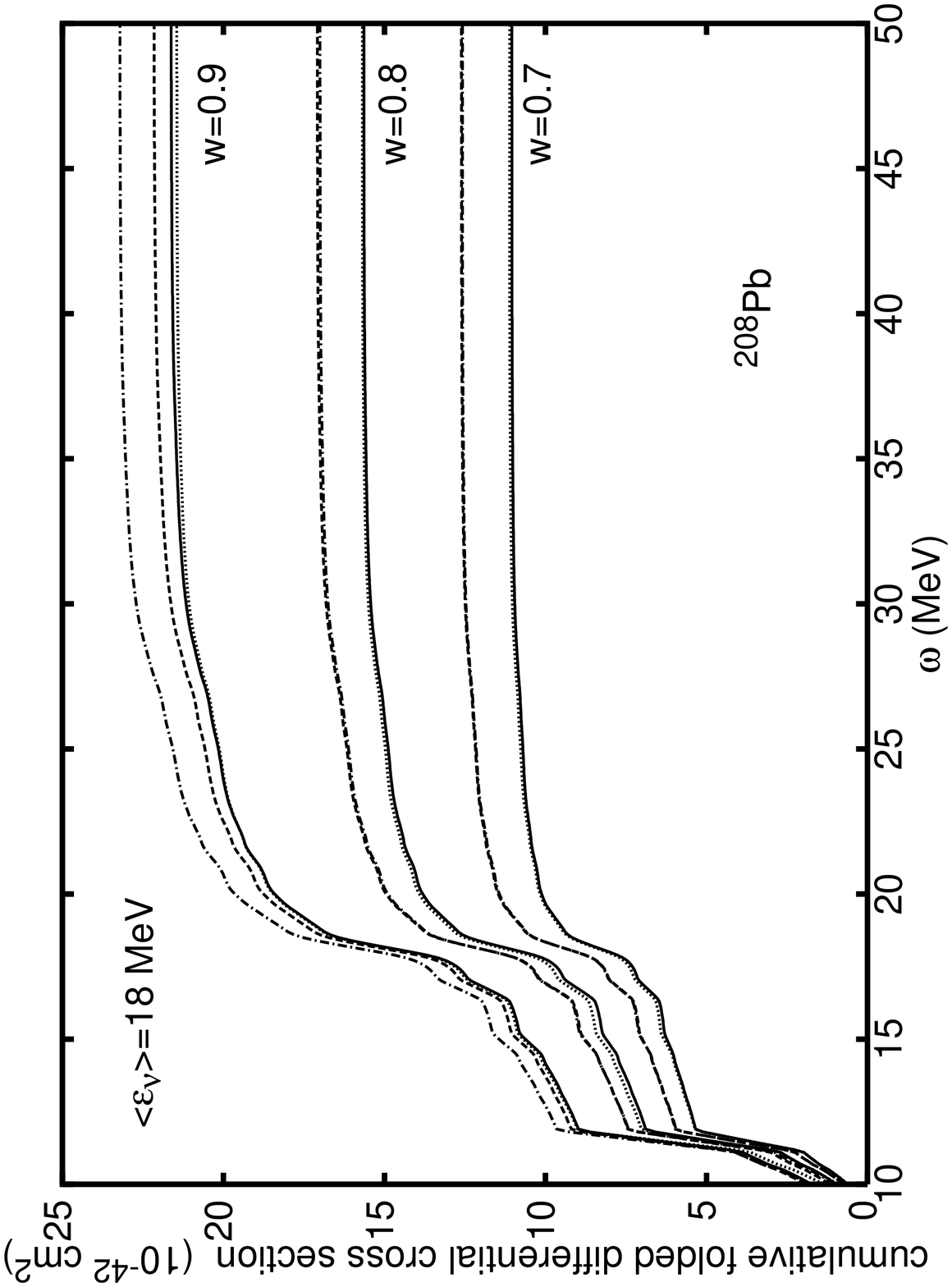"}
\caption{Cumulative differential folded cross section for scattering off $^{208}$Pb. The figure compares power-law (full line) and Fermi-Dirac spectra (dashed) with the same average energies and widths and their respective fits (dotted for the fit to the power-law spectrum, dash-dotted for the fit to the Fermi-Dirac spectra).   The parameter $w$ refers to the width relative to that of a Maxwell-Boltzmann distribution with $w$=1.}
\label{fdfig}
\end{figure}

The very satisfying agreement between original and constructed spectrum, especially for massive nuclei, suggests that it is possible
to construct supernova-neutrino signals and neutrinonucleosynthesis cross sections, simply by taking the results of
the beta-beam measurement in the detector, without going through the 
intermediate step of using the nuclear structure calculation.
The procedure is easy.
For each set of beta-beam data at a given gamma, there will be a measured 
response in the detector.  For neutral current on $^{16}$O and $^{208}$Pb it will be
spalled neutrons, protons and coincident gammas, for the charged-current 
channels it will be electrons or positrons plus possibly coincident neutrons,
protons and electrons. Our results for the differential cross sections now clearly show that taking the appropriate linear combinations of the measured response provides a very accurate picture of the neutrino-nucleosynthesis processes or of  the response of the detector to an incoming supernova-neutrino spectrum.

The accuracy reached by the fit ensures that it becomes possible to discriminate between supernova neutrino spectra that are quite similar \cite{Jachowicz:2006xx}. Fig.~\ref{fdfig} illustrates that even for  equivalent power-law and Fermi-Dirac spectra, identical up to the second moment of the distribution, the fit allows to distinguish between both responses.  The procedure is sensitive to very subtle differences in the spectrum.

In fact, as was shown in Ref.~\cite{Jachowicz:2006xx} the discriminative power of the procedure is such that it becomes possible to  reconstruct the incoming supernova neutrino spectrum from the response in a terrestrial neutrino detector.  Hence,  this technique provides a way to study the supernovaneutrinos without having to rely on  the intermediate step of theoretical cross section estimates with their related uncertainties.

This opens ways  to a model-independent  study of  oscillations in supernovaneutrinos. 
In terms of energy distributions, the supernova neutrinos are emitted with a low-energy and a high-energy component, the first associated with electron (anti)neutrinos, the latter with the heavy-flavor mu- and tau neutrinos.  In a terrestrial detector, both types will interact through neutral-current reactions. Charged-current detection channels are only accessible for electron-type neutrinos, as the production of a massive mu- or taulepton requests more energy than is available for supernova neutrinos. 
If we denote the low-energy components  by
$n^l_{SN}(\varepsilon_{{\nu}_e})$,
$n^l_{SN}(\varepsilon_{\overline{{\nu}}_e})$ and the  high-energy
components  by
$n^h_{SN}(\varepsilon_{{\nu}_{\mu}},\varepsilon_{{\nu}_{\tau}})$, and $n^h_{SN}(\varepsilon_{\overline{{\nu}}_{\mu}},\varepsilon_{\overline{{\nu}}_{\tau}})$, 
the original energy distributions can be written as 
\begin{eqnarray}
n_{SN}(\varepsilon_{\nu})&=&n^l_{SN}(\varepsilon_{\nu_e})+\,2\,n^h_{SN}(\varepsilon_{\nu_{\mu}},\varepsilon_{\nu_{\tau}}),\\
n_{SN}(\varepsilon_{\overline{\nu}})&=&n^l_{SN}(\varepsilon_{\overline{\nu}_e})+\,2\,n^h_{SN}(\varepsilon_{\overline{\nu}_\mu},\varepsilon_{\overline{\nu}_\tau}).
\end{eqnarray}

If we assume that neutrinos of all flavors are emitted in equal numbers, there will be twice as much high-energy neutrinos.   
Oscillations will transform the high and low-energy  components in a different way.  This does not affect the neutral current signal, as it is not sensitive to the neutrino flavor.   In terms of the energy spectra, the neutral current signal can then be written as
\begin{eqnarray}
S_{NC}(\omega)=\int_{\varepsilon_{\nu}} \,d{\varepsilon_{\nu}}&& \left(n^l_{SN}(\varepsilon_{\nu_e})\,+\,2\,n^h_{SN}(\varepsilon_{{\nu}_{\mu}},\varepsilon_{{\nu}_{\tau}})\right.\nonumber\\&&+\left.n^l_{SN}(\varepsilon_{\overline{\nu}_e})+\,2\,n^h_{SN}(\varepsilon_{\overline{\nu}_\mu},\varepsilon_{\overline{\nu}_\tau})\right) \nonumber\\&&\times \;\sigma_{NC}(\varepsilon_{\nu},\omega),\label{sigosc}
\end{eqnarray}
where the energy integration runs over all neutrino flavors.
The energy distribution of the neutrinos responsible for the charged current signal will be distorted according to~:
\begin{equation}
S_{CC}(\omega)=\int_{\varepsilon_{\nu}} \,d{\varepsilon_{\nu}} (R\;n^l_{SN}(\varepsilon_{\nu})\,+\,I\;n^h_{SN}(\varepsilon_{\nu})) \sigma_{CC}(\varepsilon_{\nu},\omega),\label{sigosc1}
\end{equation}
with $R$ denoting the fraction of electron-type neutrinos that remained unchanged and $I$ the fraction of heavy-flavor neutrinos that were transformed into electron neutrinos.  In the equation above $R + I = 1$, for any given energy of the neutrinos.   For simplicity, in our
example below, we assume that the neutrinos will evolve either completely
adiabatically or completely non-adiabatically so that in addition, both $R$
and $I$ are constant as a function of neutrino energy,  although our analysis could be
expanded to include the more general case.
The parameters $R$ and $I$ contain the information about oscillations that the supernovaneutrinos are carrying and can be recovered from the signal in the detector
in two steps.
First, from a set of constructed spectra $n_{N\gamma}$, the two linear combinations of beta-beam responses 
\begin{equation}
n^{l,fit}_{N\gamma}(\varepsilon_{\nu})=\sum_{i=1}^N a_i^{l,fit} n_{\gamma_i}(\varepsilon_{\nu}),
\end{equation}
and \begin{equation}
n^{h,fit}_{N\gamma}(\varepsilon_{\nu})=\sum_{i=1}^N a_i^{h,fit} n_{\gamma_i}(\varepsilon_{\nu})
\end{equation}
for which
\begin{equation}
2\,\int_{\varepsilon_{\nu}} \,d{\varepsilon_{\nu}} (n^{l,fit}(\varepsilon_{\nu})\,+\,2\,n^{h,fit}(\varepsilon_{\nu})) \sigma_{NC}(\varepsilon_{\nu}),
\label{fitosc1}\end{equation}
minimizes the difference with the neutral current detector signal from Eq.~\ref{sigosc}
 are selected. 
 In the neutral-current signal,  we have not distinguished
the antineutrino component from the neutrino component.
The antineutrino signal is not too different
 from its neutrino counterpart, and so we have disregarded these differences
 in this illustration of the proposed reconstruction method.
The resulting  two sets of expansion parameters $(a_i^{fit,l}\,;\,i=3,15)$ and $(a_i^{fit,h}\,;\,i=3,15)$ are then used to construct combinations 
\begin{equation}
\int_{\varepsilon_{\nu}} \,d{\varepsilon_{\nu}} (R^{fit}\;n^{l,fit}(\varepsilon_{\nu})\,+\,I^{fit}\;n^{h,fit}(\varepsilon_{\nu})) \sigma_{CC}(\varepsilon_{\nu}),\label{fitosc}
\end{equation}
and seek for the oscillation parameters $R^{fit}$ and $I^{fit}$ that yield the best agreement between the expression of Eq.~\ref{fitosc} and the signal of Eq.~\ref{sigosc1}.  

\begin{figure*}[hbt]
\vspace*{12.5cm}
\special{hscale=35 vscale=35 hsize=1500 vsize=600
         hoffset=-15 voffset=380 angle=-90 psfile="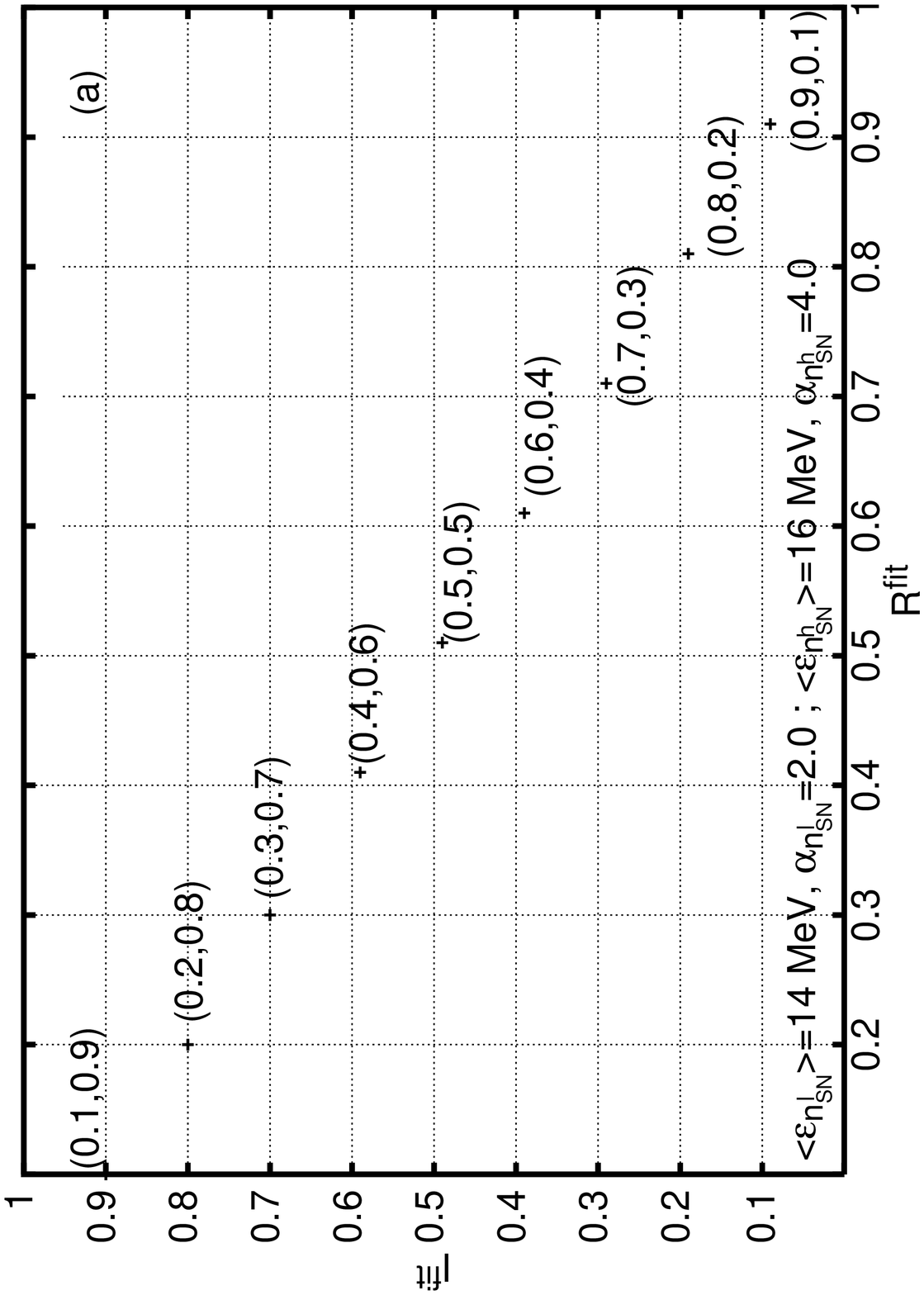"}
\special{hscale=35 vscale=35 hsize=1500 vsize=600
         hoffset=235 voffset=380 angle=-90 psfile="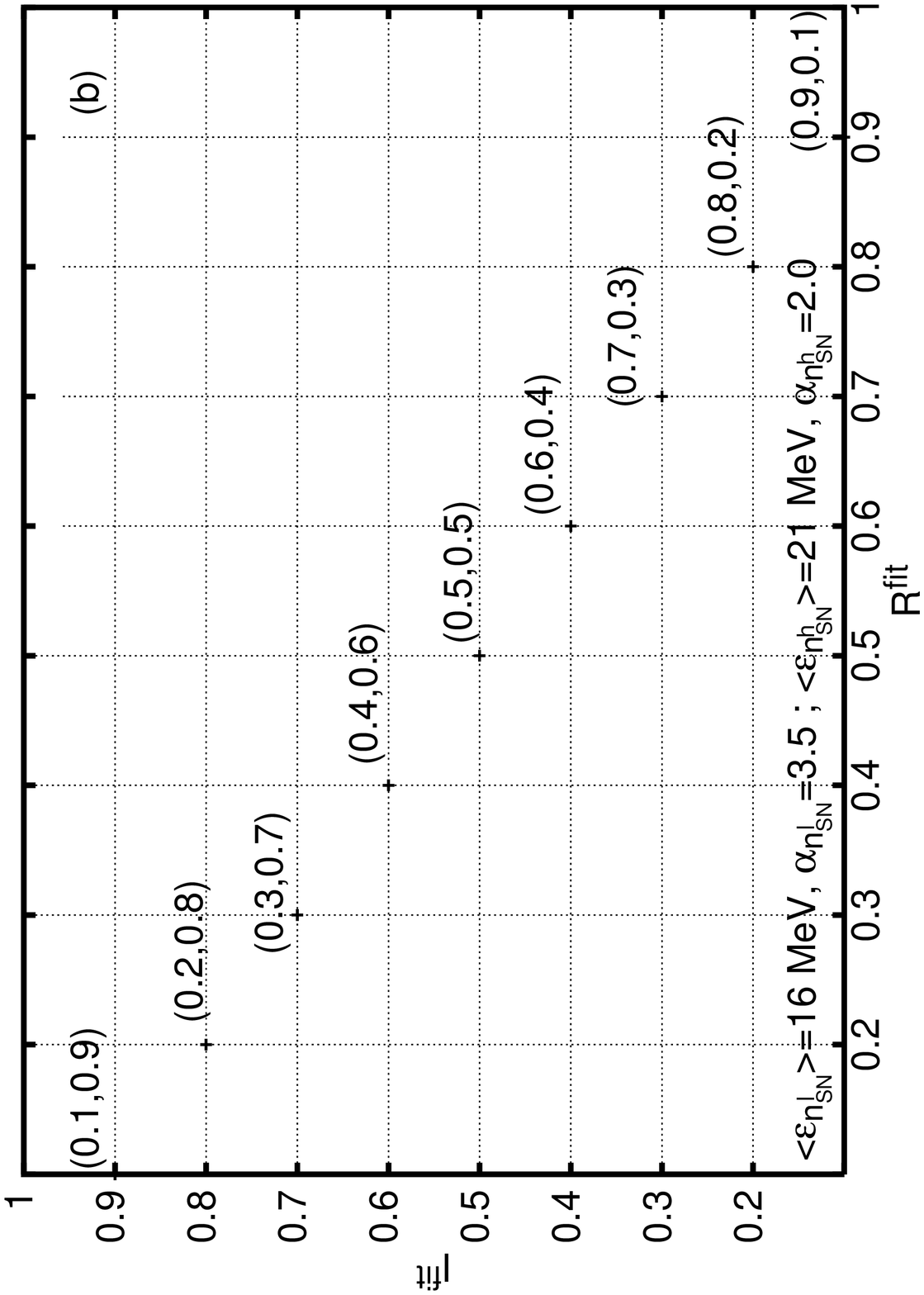"}
\special{hscale=35 vscale=35 hsize=1500 vsize=600
         hoffset=-15 voffset=200 angle=-90 psfile="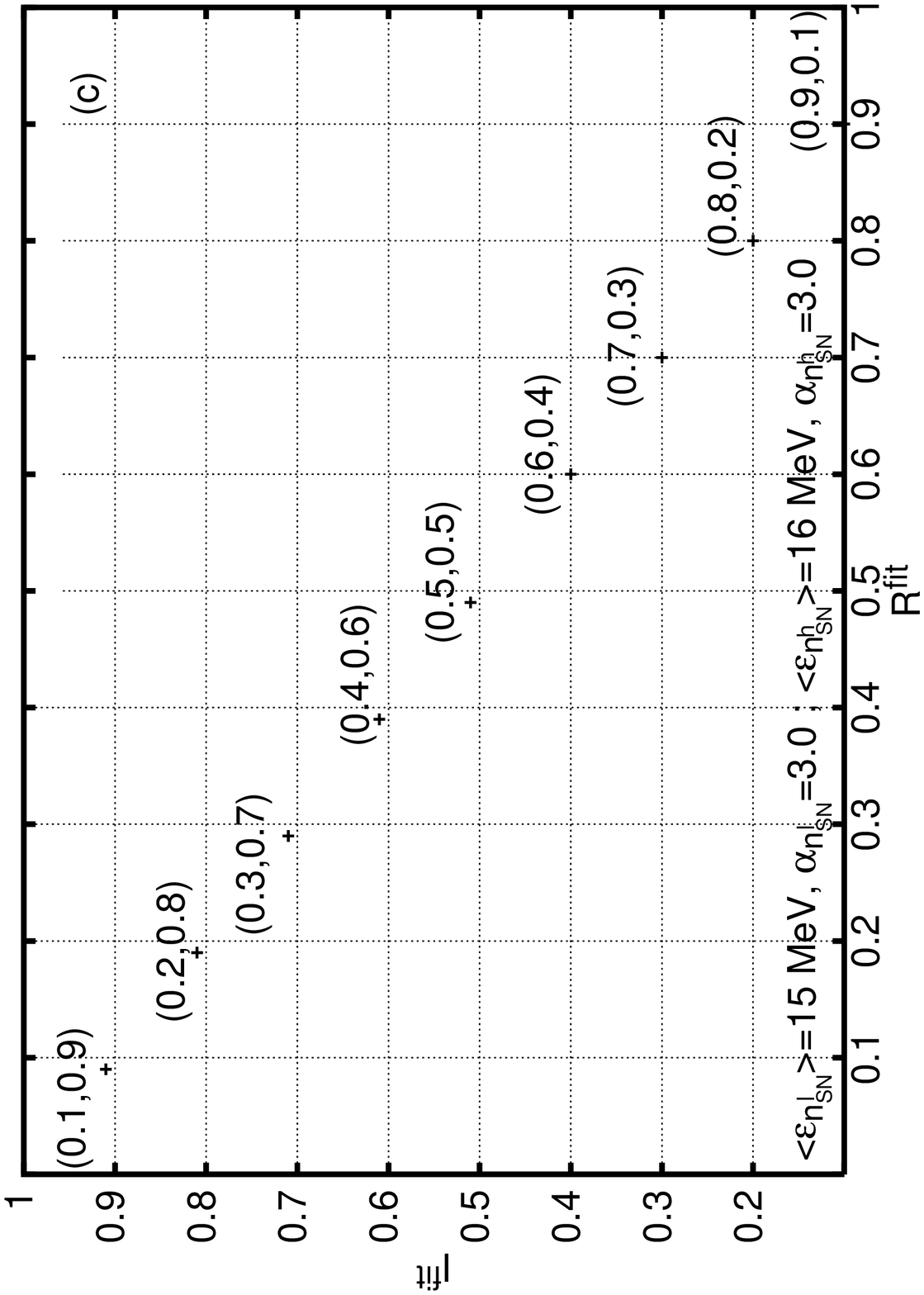"}
\special{hscale=35 vscale=35 hsize=1500 vsize=600
         hoffset=235 voffset=200 angle=-90 psfile="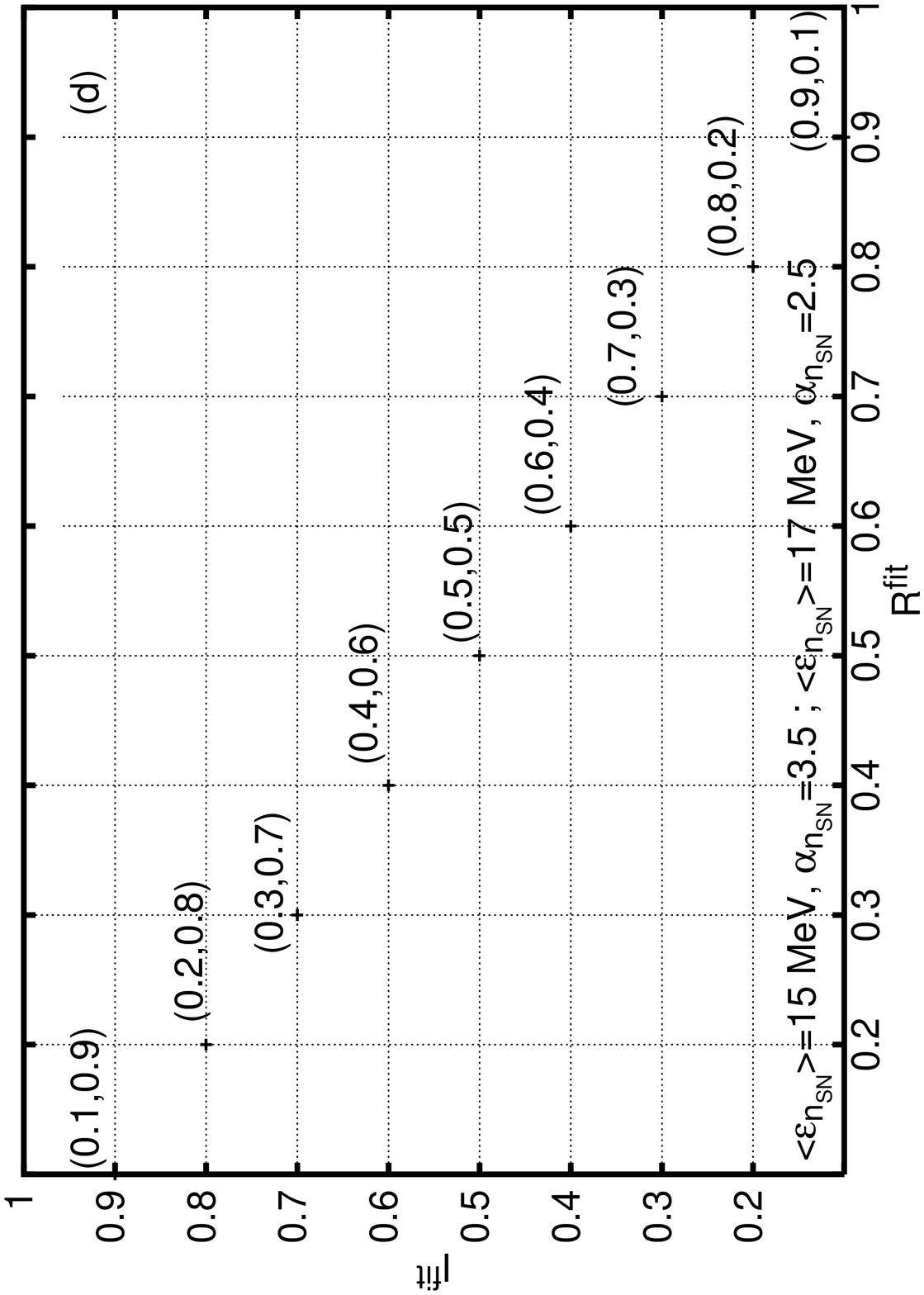"}
\caption{Extraction of the oscillation parameters $R^{fit}$ and $I^{fit}$ from a fit of the expressions in Eqs.~(\ref{fitosc1}) and (\ref{fitosc}) to the neutral $S_{NC}(\omega)=\int_{\varepsilon_{\nu}} \,d{\varepsilon_{\nu}} (n^l_{SN}(\varepsilon_{\nu})+\,2\,n^h_{SN}(\varepsilon_{\nu})) \sigma_{NC}(\varepsilon_{\nu},\omega)$, and charged-current $S_{CC}(\omega)=\int_{\varepsilon_{\nu}} \,d{\varepsilon_{\nu}} (R\;n^l_{SN}(\varepsilon_{\nu})+\,I\;n^h_{SN}(\varepsilon_{\nu})) \sigma_{CC}(\varepsilon_{\nu},\omega)$ supernova-neutrino signal in an oxygen detector.   For each fitted point  the original parameter values $(R,I)$  are given between brackets.  The agreement between ($R$, $I$) and the reconstructed ($R^{fit}$, $I^{fit}$) is very good.}
\label{oscfig}
\end{figure*}

\begin{figure}
\vspace*{6.5cm}
\special{hscale=36 vscale=36 hsize=1500 vsize=600
         hoffset=-20 voffset=200 angle=-90 psfile="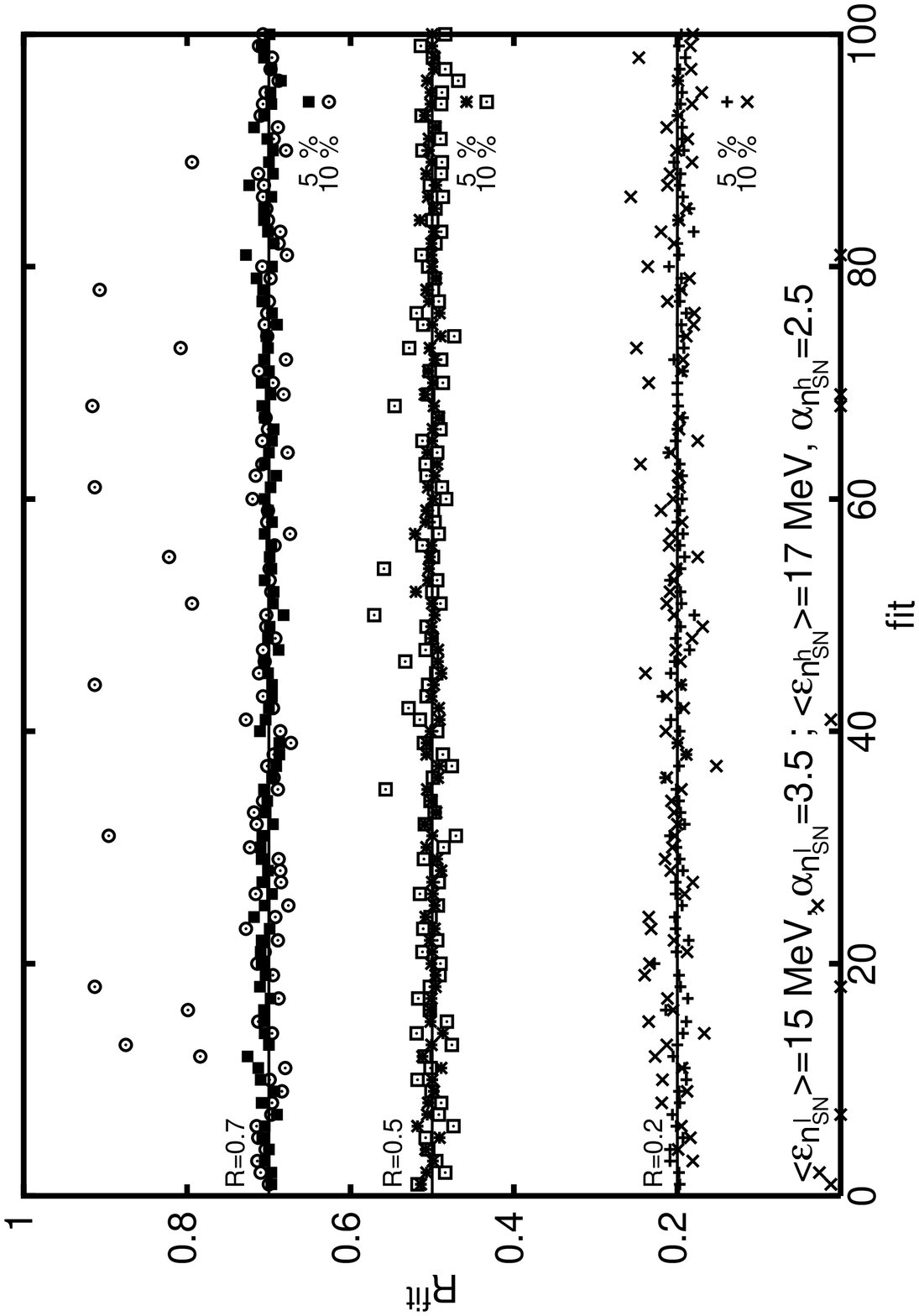"}
\caption{Influence of noise on the reconstruction of the parameter $R$ from the supernova neutrino signal in a terrestrial detector.  The figure shows the value of $R^{fit}$, for original values $R$=0.2, 0.5 and 0.7  for 100 fits to a signal subject to random noise at 5 and 10\% level, as described in the text.} 
\label{noisfig}
\end{figure} 

The results of this procedure are shown in Fig.~\ref{oscfig}, where the parameters $R$ and $I$ were extracted from a supernova-neutrino signal consisting of two mixed power-law distributions.  The reconstruction was realized using a library of 125 spectra $n^{fit}_{\gamma=3,...,15}(\varepsilon_{\nu})$ with average energies between 12 and 24 MeV and width varying from $\alpha=$ 2 to 5.
 The method is very successful, even when the spectra are very close together as illustrated in panel (c) of the figure.

As for most neutrino experiments, especially at relatively low energies, statistics can be expected to be rather poor.  This will result in uncertainties on the observed supernova-neutrino signal and on the beta-beam data.  We investigated the influence of these uncertainties on the reconstruction procedure by adding random noise to the 'data points' of the signal of Eqs.~\ref{sigosc} and \ref{sigosc1}
  i.e $S_{NC}(\omega)$ and $S_{CC}(\omega)$ were replaced by $S(\omega)*(1.+p*r)$, where $r$ was chosen radomly between 0 and 1 for each data point and $p$ determines the size of the noise, $p=0.05$ and $p=0.10$ for the simulations shown in  Fig.~\ref{noisfig} .
The accuracy of the first step in the procedure , i.e. the reconstruction of the spectra $n^l_{SN}(\varepsilon_{\nu})$ and $n^l_{SN}(\varepsilon_{\nu})$ from the neutral current signal $S_{NC}(\omega)$
is of course crucial for the subsequent determination of $R$ and $I$.
It was already shown in Ref.~\cite{Jachowicz:2006xx} that the procedure for reconstructing a single spectrum is stable against uncertainties at this level. 
Fig.~\ref{noisfig} corroborates these results and shows that the second step, the reconstruction of the parameters $R$ and $I$ from the fit to the charged current spectrum is also  stable against these uncertainties.  For a fit to 100 different spectra, all obtained by adding random noise to the signal, the results for $R^{fit}$ are strongly concentrated around the original $R$-values. In agreement with the results of Ref.~\cite{Jachowicz:2006xx}, the reconstructive power of the proposed procedure slightly dwindles for spectra at lower energies where the influence of uncertainties becomes larger.

Concluding, we have shown that beta-beam measurements can be of great help in the analysis of a supernova-neutrino signal.  Making use of the information provided by beta-beam data provides a model-independent way  for the extraction of  information on the supernova-neutrino energy-distribution from its signal in a terrestrial detector. As neutrinos are the only particles reaching us from the core of the exploding star, this makes beta-beams an important  instrument for probing  the processes driving the core-collapse supernova and the circumstances reigning in the heart of the event. The central idea of the proposed technique consists of a reconstruction of supernova-neutrino responses using experimental beta-beam data, hence it is also of direct use for the prediction of neutrinonucleosythesis reaction products.
We  propose to use this technique as a lever to make advantage of beta-beam data  in disentangling the oscillation characteristics  from mixed spectra.  We have  shown that the procedure is effective in extracting oscillation parameters from the signal in a terrestrial detector, illustrating the use of the technique 
for  gathering information about the supernova as well as its neutrinos.

\acknowledgements
The authors would like to thank M.~Lindroos,  and K.~Heyde for interesting discussions. N.J.~thanks the Research Foundation Flanders (FWO) for financial support. G.C.M.~acknowledges support from the Department of Energy, under contract DE-FG02-02ER41216. C.V. acknowledges the financial support of the EC under the FP6  'Research Infrastructure Action - Structuring the European Research Area' EURISOL DS Project, Contract Number 515768 RIDS.

\end{document}